\documentclass[aps,twocolumn,pra,groupedaddress,superscriptaddress,nofootinbib]{revtex4}

\usepackage{graphicx,amsmath,amssymb,amsbsy,subfigure,hyperref,bbm,times,txfonts,float}
\usepackage{subfigure,hyperref,bbm,times}
\usepackage[T1]{fontenc}
\usepackage{braket}
\usepackage{epsfig}
\usepackage{color}
\usepackage{graphicx}
\usepackage{dcolumn}
\usepackage{bm}

\providecommand{\openone}{\leavevmode\hbox{\small1\kern-3.8pt\normalsize1}}

\usepackage{soul}

\hypersetup{
	colorlinks=true,
	linkcolor=blue,
}
\graphicspath{{figure/}}

\begin{document}
	
	
	\title{Relativistic quantum thermometry through a moving sensor}

\author{Hossein Rangani Jahromi}
\email{h.ranganijahromi@jahromu.ac.ir}%
\affiliation{Physics Department, Faculty of Sciences, Jahrom University, P.B. 74135111, Jahrom, Iran}

\author{Samira Ebrahimi Asl Mamaghani}%
\affiliation{Dipartimento di Ingegneria, Universit\`{a} di Palermo, Viale delle Scienze, 90128 Palermo, Italy}%
\affiliation{INRS-EMT, 1650 Boulevard Lionel-Boulet, Varennes, Qu\'{e}bec J3X 1S2, Canada}

\author{Rosario Lo Franco}%
\email{rosario.lofranco@unipa.it}
\affiliation{Dipartimento di Ingegneria, Universit\`{a} di Palermo, Viale delle Scienze, 90128 Palermo, Italy}%

\date{\today}

\begin{abstract}
Using a two-level moving probe, we address the temperature estimation of a static thermal bath modeled by a massless scalar field prepared in a thermal state. Different couplings of the probe to the field are discussed under various scenarios. We find that the thermometry is completely unaffected by the Lamb shift of the energy levels. We take into account the roles of probe velocity, its initial preparation, and environmental control parameters for achieving optimal temperature estimation. We show that a practical technique can be utilized to implement such a quantum thermometry. Finally, exploiting the thermal sensor moving at high velocity to probe temperature within a multiparameter-estimation strategy, we demonstrate perfect supremacy of the joint estimation over the individual one.
\end{abstract}


\maketitle


\section{Introduction \label{introduction}}
 Temperature estimation is an important task on all scales, ranging from atomic systems  near absolute zero to astronomical bodies of high temperatures. Particularly, applications of thermometry in microscale and nanoscale 
devices are becoming  in great demand  as technology
advances \cite{yue2012nanoscale,giazotto2006opportunities,carlos2015thermometry,mehboudi2019thermometry}. Examples  include precise  temperature
estimation of ultracold gases \cite{sabin2014impurities,mehboudi2019using,bouton2020single}, electrons in superconductors \cite{gasparinetti2015fast,halbertal2016nanoscale}, and the application of atomic-size devices such as quantum dots or color centers in diamond, when probes are used in various systems  \cite{kucsko2013nanometre,neumann2013high,haupt2014single,hovhannisyan2021optimal}.
The classical approach to thermometry is that the state of thermometer, brought into thermal contact with a sample, is monitored
for some time, conveying the information about the sample temperature \cite{jorgensen2020tight}.

In the quantum realm, thermodynamical quantities are usually
challenging to define, manipulate and measure  \cite{binder2018thermodynamics}, to the point that they may  lead to reformulation of the thermodynamics laws \cite{levy2012quantum,kolavr2012quantum,brandao2015second,uzdin2018global,mehboudi2019thermometry}. Recent advancements in quantum metrology \cite{helstrom1969quantum,paris2009quantum,giovannetti2006quantum,giovannetti2011advances,degen2017quantum,pirandola2018advances,pezze2018quantum,huang2019cryptographic,liu2022optimal,jahromi2021hilbert,jahromi2018parameter} have led to extension of
 thermodynamics boundaries into novel territories, in which tiny objects are cooled to ultra-low temperatures \cite{giazotto2006opportunities,bloch2008many}, and  resulted in the development of a fast-growing field of research, i.e.,
 \textit{quantum thermometry} \cite{goold2016role,stace2010quantum,seah2019collisional,montenegro2020mechanical,zhang2021non,mok2021optimal,mancino2020nonequilibrium,rubio2021global,paris2015achieving,pasquale2018quantum,correa2017enhancement,planella2022bath,hovhannisyan2021optimal,alves2022bayesian,jorgensen2020tight,gebbia2020two,luiz2022machine,o2021stochastic,kenfack2021quantum,farajollahi2018estimation,jahromi2020quantum}. The basic idea is  to estimate the temperature $ T $ of a thermal environment  by
letting it  interact with a quantum system, say a
qubit or a pair of entangled qubits, called a quantum \textit{probe}, which
 are then subsequently measured to extract the information. Applying quantum probes  to estimate parameters of interest has  the advantage that it does not
 perturb too much the system under investigation. Provided that the probe reaches a   non-equilibrium steady state, or
 thermal equilibrium with the sample,  the optimal measurement, minimizing
 the uncertainty of the thermometry through saturating the Cram\'{e}r-Rao inequality \cite{paris2009quantum}
 may be achievable \cite{correa2015individual,correa2017enhancement,mehboudi2019using,hofer2017quantum,guarnieri2019thermodynamics,jorgensen2020tight}. In the non-equilibrium dynamics where the temperature is extracted  from the state
 of the probe before its thermalization \cite{montenegro2020mechanical,brunelli2012qubit,de2017estimating,feyles2019dynamical,gebbia2020two,seah2019collisional,mancino2020nonequilibrium}, the optimal thermometry,  generally depends on the unknown temperature
 of the system, making its achievement challenging in practice \cite{mok2021optimal}.
 
 \par
 
In  near-equilibrium thermometry, it is known that the
energy measurement is  the optimal choice \cite{correa2015individual,marzolino2013precision,hovhannisyan2021optimal},  as happens in classical
physics. Nonetheless,  this approach may be very demanding, because
it requires access to the total system, measurement of its energy
and   full knowledge of the spectrum. However, when a small quantum probe interacts with the system
without causing much disturbance and then it is  measured, those limitations are not encountered \cite{montenegro2020mechanical,mok2021optimal}. These considerations motivate more investigation of quantum metrology in the context of near-equilibrium thermometry, especially for  relativistic scenarios in which the observed temperature by the moving system is a challenging discussion \cite{papadatos2020relativistic,cai2021velocity}.

 \par
Quantum field theory indicates that  different noninertial observers do not agrees on the number of particles in a given field state. In fact,  a
quantum (scalar) field in the Minkowski-Unruh vacuum
from an inertial perspective is observed as a thermal bath by a
Rindler observer moving with uniform acceleration; this phenomenon is known as \textit{Unruh effect} \cite{crispino2008unruh,schlicht2004considerations}. Quantum estimation of the Unruh temperature by various accelerated  probes have been investigated in some references \cite{aspachs2010optimal,tian2015relativistic,wang2014quantum,feng2022quantum,patterson2022fisher}. Moreover, in the absence of Unruh effect,  the  temperature estimation through a static atom immersed in a thermal bath with a boundary  in
a massless scalar field has been analyzed in \cite{zhao2020quantum}. In addition,  the performance of estimating
the parameters encoded into the initial state of  a two-level
probe, moving with a constant velocity, as the detector coupled to a massless scalar field, has been also studied \cite{liu2021does}. It has been
shown that the estimation is
completely unaffected by the velocity, however, it  becomes more inaccurate over time because of the decoherence caused by
the interaction between the atom and the field. Hence, it would be interesting to investigate how quantum thermometry  is affected in this scenario.
\par
In this paper, we consider an Unruh-DeWitt (UDW) detector \cite{costa1995temperature} moving with constant velocity and interacting with a thermal scalar field. Because of this interaction, the UDW detector \cite{unruh1976notes,hu2012relativistic} is an open  quantum system \cite{moustos2018asymptotic,zhao2020quantum,papadatos2020relativistic,liu2021does,ng2018unruh,jafarzadeh2018teleportation,rangani2019weak}  encoding the information on the temperature of the field, and hence playing the role of a quantum thermometer. Employing a moving probe for quantum estimation is a powerful technique, especially when the metrological setup used to analyze and extract information from the probe is located elsewhere. Here, after computing quantum Fisher information, a reliability measure of the moving probe as a temperature sensor, we investigate how the initial parameters, as well as the ambient ones, can be controlled to improve the thermometry. In particular, we find that control over the probe velocity plays a key role to achieve optimal accuracy. We also elaborate on a physical proposition for the experimental implementation of the quantum thermometry. Finally, we discuss how the thermometer can be used for the simultaneous estimation of parameters and illustrate the perfect supremacy of quantum thermometry in a multiparameter-estimation scenario. 
\par
The paper is structured as follows: In Sec. \ref{QuantumFI}, a brief review of the theory of quantum metrology is presented. Then, we introduce the model in Sec. \ref{Model}. The process of quantum thermometry is completely investigated in Sec. \ref{Quantum Thermometry}. Finally,  Sec. \ref{Conclusion} is devoted to summarizing and discussing the most important results. Throughout this paper, we apply units with $ c=\hbar= k_{B}=1$. Moreover, 
a simple set of scaled units is introduced \cite{jahromi2015precision} in Table I.

\begin{table}[ht]\label{T1}
	\caption{ Scaled and SI units used in this paper } 
	\centering 
	\begin{tabular}{c c c} 
		\hline\hline 
		$ \text{Physical~quantity} $ & $ \text{Scaled~unit} $ & $ \text{SI~unit} $ \\ [1ex] 
		\hline 
		Temperature: $ T $ & 1 & $ \tilde{T}=1 K $ \\ 
		Angular Frequency: $\omega $ & 1 & $ \tilde{\omega}=k_{B}\tilde{T} /\hbar$  \\
		Time:$  t $ & 1 & $ \tilde{t}=1/\tilde{\omega}$\\
		Coupling constant: $ \lambda$ & 1 & $ \tilde{ \lambda}=1/ \tilde{t}$  \\ [1ex] 
		\hline 
	\end{tabular}
	\label{table:nonlin} 
\end{table}

 \section {Quantum Fisher information}\label{QuantumFI}
 Quantum Fisher information \cite{braunstein1994statistical}, determining the fundamental
 limit to the accuracy of estimating an unknown parameter,
 plays the most key role in quantum metrology.
 First we  concisely review the principles of classical
 estimation theory and introduce the tools which  it provides to calculate  the  bounds
 to accuracy of any quantum metrology process.
 In an estimation problem we aim at   extracting the value of a
 parameter $ \lambda $ through measuring a related quantity $ X $. For solving this  problem, one should obtain an estimator $ \hat{\lambda}\equiv \hat{\lambda} (x_{1},x_{2},...)$, generating an estimate
 $ \hat{\lambda} $ for the parameter $ \lambda $, based on the achieved measurement  outcomes $ \left\{x_{k} \right\} $. In the classical theory, the variance  $ \text{Var}(\lambda)=E[\hat{\lambda}^{2}] -E[\hat{\lambda}]^{2}$
 of any unbiased estimator, in which  $ E[...] $ denotes the mean with respect to the $ n $ identically
 distributed random variables $ x_{i} $,  fulfills the
 Cramer-Rao inequality $ \text{Var}(\lambda) \geq \dfrac{1}{MF_{\lambda}} $ where $ M  $  indicates the  number of independent measurements. This  inequality  gives  a lower bound on the variance in terms of the
  the Fisher information (FI) $ F(\lambda) $
 \begin{equation}\label{cfi}
 	F_{\lambda}=\sum_{x}\dfrac{[\partial_{\lambda}p(x|\lambda)]^{2}}{p(x|\lambda)},
 \end{equation}
in which $ p(x|\lambda) $ signifies the conditional probability of obtaining the
 value $ x $ as the unknown parameter has the value $ \lambda $.   When  the eigenvalue spectrum of observable $ X $  is
 continuous, the summation in Eq. (\ref{cfi}) must be replaced by
 an integral.

 \par In the quantum regime $ p(x|\lambda)=\text{Tr}\left[ \rho P_{x}\right] $ in which  $ \rho $ indicates the state of the quantum system and $ P_{x} $ represents the probability operator-valued measure (POVM) characterizing the measurement.  In brief summary,  it is feasible to indirectly achieve the value of the
 physical parameter, intending to estimate it,  through measuring an observable $ X $ and then 
 making statistical analysis on the measurement outcomes.
 An  estimator is called efficient when it at least asymptotically saturates  the
 Cramer-Rao bound.

Obviously, various observables result in  various
 probability distributions, giving rise  to miscellaneous FIs and therefore
 to different precision for estimation of $ \lambda $.  The ultimate
 bound to the precision, determined by  the quantum Fisher information (QFI), is attained by maximizing the FI over
 the set of the observables. 
 The QFI of an unknown parameter $ \lambda $ encoded into the quantum state $ \rho\left(\lambda \right) $ is given by \cite{helstrom1969quantum,braunstein1994statistical,paris2009quantum}
 \begin{equation}\label{01}
 	H_{\lambda}=\text{Tr}\left[\rho\left(\lambda \right)L_{\lambda}^{2} \right]=\text{Tr}\left[\left( \partial_{\lambda}\rho\left(\lambda \right)\right) L_{\lambda}\right], 
 \end{equation}
where $ L_{\lambda} $ indicates the corresponding symmetric logarithmic derivative (SLD) given by $ \partial_{\lambda}\rho\left(\lambda \right)=\frac{1}{2}\left(L_{\lambda}\rho\left(\lambda \right)+\rho\left(\lambda \right)L_{\lambda}\right), $ in which $ \partial_{\lambda}=\partial/\partial\lambda $. As is well-known, the set of projectors over the eigenvectors of the SLD gives
an optimal POVM. 
\par An explicit  expression of the QFI can be obtained for single-qubit systems. It is known that any qubit state can be written in the Bloch sphere representation as $\rho=\frac{1}{2}\left(I+\boldsymbol{\omega}\cdot\boldsymbol{\sigma}\right)$,
in which $ \boldsymbol{\omega}=\left(\omega_{x},\omega_{y},\omega_{z} \right)^{T} $ denotes the Bloch vector and $ \boldsymbol{\sigma}=\left(\sigma_{x},\sigma_{y},\sigma_{z} \right)  $ represents the Pauli matrice, leading to the  following compact formula for  the QFI of the single-qubit state \cite{zhong2013fisher}

\begin{equation}\label{a12}
	H_{\lambda}=\left\{\begin{array}{cc}
		|\partial_{\lambda}\boldsymbol{\omega}|^{2}+\frac{\left( \boldsymbol{\omega}\cdot\partial_{\lambda}\boldsymbol{\omega}\right) ^{2}}{1-|\boldsymbol{\omega}|^{2}},&~~~~~~~~~~ |\boldsymbol{\omega}|<1, \\
		|\partial_{\lambda}\boldsymbol{\omega}|^{2},~~~~~~~~~~~~~~~~~~~ & ~~~~~~~~~~|\boldsymbol{\omega}|=1. \\
	\end{array}\right.
\end{equation}
where $ |\boldsymbol{\omega}|<1 ~(|\boldsymbol{\omega}|=1)$ is used for a mixed (pure) state.

\section{Model}\label{Model}

We consider a composite quantum system consisting of
a microscopic two-level probe S, for example, an atom or a molecule, and a quantum scalar field $ \hat{\Phi}(x) $ \cite{papadatos2020relativistic}. The
system is characterized by a Hilbert space $ \mathcal{H}_{S}\otimes \mathcal{H}_{\Phi} $, in which $ \mathcal{H}_{S}$ ($ \mathcal{H}_{\Phi} $)
denotes the Hilbert space of the probe (field).
The Hamiltonian is given by
\begin{equation}\label{Hamiltonian}
	\hat{H}=\hat{h}\otimes \hat{I} +\hat{I} \otimes \hat{H}_{\Phi}+\hat{V},
\end{equation}
 where $\hat{h}$ represents the Hamiltonian generating time translations with respect to  proper time parameter $\tau$ of $S$.
 The Hamiltonian $  \hat{H}_{\Phi} $, describing a free massless scalar field, is given by  $\hat{H}_{\Phi}=\frac{1}{2}\int d^{3}x({{\hat{\pi}}^{2}}+{(\nabla \hat{\Phi})^{2}})$ in which $\hat{\pi}(x)$ denotes the conjugate momentum of the field $\hat{\Phi}(x)$. The general form of the  interaction terms is $\hat{V}=\lambda \hat{A}\otimes \hat{O}(x(\tau))$ where $ \lambda $ represents a coupling constant, $ \hat{A} $ denotes a self-adjoint operator
 on $ \mathcal{H}_{S} $, and $ \hat{O} $ designates a local
 composite operator for the scalar field. Moreover, the  trajectory $x=(t,\textbf{x})$, in which  $ \textbf{x}(\tau) $ represents the detector path,  is given by 
  
  \begin{equation}
  	x(\tau)=(\cosh{u},\sinh{u},0,0)\tau
  \end{equation}
 where $ u $ represents the rapidity of the trajectory,  associated with velocity $v=\tanh{u}$, and $\lambda$ denotes a coupling constant.
 \par
We assume a seperable initial state $ \hat{\rho}_{0} \otimes  \hat{\rho}_{\Phi}$ where $  \hat{\rho}_{\Phi}$ represents a Gibbs state at temperature $ \beta^{-1} $, $ \hat{\rho}_{\Phi}=\text{e}^{-\beta \mathcal{H}_{\Phi}}/\text{Tr}\big[\text{e}^{-\beta \mathcal{H}_{\Phi}}\big] $. We focus on two different coupling regimes, i.e., (i)  $  \hat{O}(x(\tau))=\hat{\Phi}(x)  $ and (ii)  $  \hat{O}(x(\tau))=\dot{\hat{\Phi}}(x)  $ representing, respectively, the Unruh-DeWitt (UDW) and time-derivative (TD) couplings.
 
 \par
 Starting from the Hamiltonian (\ref{Hamiltonian}) with the separable initial states and defining  the \textit{transition operators}
 \begin{equation}
 	\hat{A}_{\omega}=\sum_{n,m,\epsilon_{m}-\epsilon_{n}=\omega} \Bra{n}\hat{A}\ket{m}\ket{n}\bra{m},
 \end{equation}
 one can extract the 
 the reduced dynamics of the probe  from 
 the following \textit{second-order master equation}
 
 \begin{equation}
 	\frac{\partial{\hat{\rho}}}{\partial{{\tau}}}=-[\hat{h}+\hat{h}_{LS},\hat{\rho}]+\sum_{\omega}\Gamma(\omega)[\hat{A}_{\omega}\hat{\rho}\hat{A}^\dagger_{\omega} - \frac{1}{2}\hat{A}{^{\dagger}}_{\omega}\hat{A}_{\omega}\hat{\rho}-\frac{1}{2}\hat{\rho}\hat{A}{^{\dagger}}_{\omega}\hat{A}_{\omega}],
 	\end{equation}
where $\omega=\epsilon_{m}-\epsilon_{n}$ denotes the set of all energy differences. Moreover, 
$\hat{h}_{LS}=\sum_{\omega}\Delta{(\omega)}\hat{A}^\dagger_{\omega} \hat{A}_{\omega}$ in which $\Delta{(\omega)}$ represents the Lamb shift  of the energy levels, and
\begin{equation}
	\Gamma(\omega)=\gamma(|\omega|)
	\begin{cases}
		1+N(|\omega|), & { \omega>0}\\
		N(|\omega|), & { \omega<0}.
	\end{cases}
\end{equation}
For the UDW coupling with positive $\omega  $,  the  expressions for $\gamma(\omega)$  and $N(\omega)$ are given by 
 \begin{equation}
 	\gamma_{{UDW}}(\omega)=\lambda^2/2\pi \omega,
 \end{equation}
 \begin{equation}
 	N_{{UDW}}(\omega){=} \frac{1}{2 \omega \beta \sinh (u)} \log \frac{1-e^{-\beta \omega e^ {u}}}{1-e^{ -\beta \omega e^{ (-u)}}}
 \end{equation}
where  $ n_{k}=(\text{e}^{\beta k}-1)^{-1}$ in which $ k=|\textbf{k}| $, denotes the expected numbers of particles of momentum $ \textbf{k} $. Moreover, 
 \begin{equation}
 	\Delta _{{UDW}}(\omega){=}\left[{sgn}\left(\omega\right) \left(\Delta _0+\frac{\lambda ^2 \left(\int_0^{\infty } n_k \log \left(\frac{\left(k e^{-u}+\omega\right) \left(\omega-k e^{u}\right)}{\left(\omega-k e^{-u}\right) \left(k e^{u}+\omega\right)}\right) \, dk\right)}{8 \pi ^2 \sinh (u)}\right)\right]
 \end{equation}
where
 \begin{equation}\label{delta0}
 	\Delta _0{=}\frac{\left(\lambda ^2 \left| \omega\right| \right) \log \left(\epsilon  e^{ (\gamma -1)} \left| \omega\right| \right)}{4 \pi ^2}, 
 \end{equation} 
indicates the contribution for the Lamb-Shift. In (\ref{delta0})  $ \epsilon $ and $ \gamma $ signify, respectively, the cutoff and the Euler-Macheronni constant.
\par 
The corresponding expressions for $ \gamma(\omega) $, $ N(\omega) $, and $ 	\Delta _{{TD}}(\omega) $  in the TD-coupling regime are
 \begin{equation}
 	\gamma _{{TD}}(\omega){=}\frac{ \left(\lambda ^2 (2 \cosh (2 u)+1)\right)}{6 \pi }\omega ^3,
 \end{equation}
 \begin{equation}
 	N_{{TD}}(\omega){=}\frac{3 \int_{\omega e^{-u}}^{\omega  e^{u}} k^2 n_k \, dk}{2 \omega ^3 \sinh (u) (2 \cosh (2 u)+1)},
 \end{equation}
and
 \begin{equation}
 	\Delta _{{TD}}(\omega)=\left[{\text{sgn}}\left(\omega\right) \left(\frac{\lambda ^2 \left(\int_0^{\infty } k^2 n_k \log \left(\frac{\left(k e^{-u}+\omega\right) \left(\omega-k e^{u}\right)}{\left(\omega-k e^{-u}\right) \left(k e^{u}+\omega\right)}\right)\, dk\right)}{8 \pi ^2 \sinh (u)}+\tilde{\Delta} _{0}\right)\right],
 \end{equation}
where 
 \begin{equation}
 	\tilde{\Delta} _{0}{=}\frac{\lambda ^2 (2 \cosh (2 u)+1) \left| \omega \right| {}^3}{12 \pi ^2}\bigg(\dfrac{3}{(\omega  \epsilon)^{2}}+\text{log}(|\omega| \epsilon \text{e}^{\gamma-1})\bigg).
 \end{equation}
 
 We assume the Hamiltonian of the two-level atom characterized by frequency $\Omega_0$  is given by $h=\frac{1}{2}$ $\Omega_0$ $\sigma_z$ and the coupling operator is expressed as $\hat{A}=\sigma_1$, resulting in two transition operators $\hat{A}_{\Omega_0}=\sigma_{-}$ and $\hat{A}_{-\Omega_0}=\sigma_{+}$. Solving the corresponding master equation  for the initial pure state 
 \begin{equation}
 	\ket{\psi_{0}}=e^{i\phi} \cos (\frac{\theta}{2})\ket{1}+\sin(\frac{\theta}{2})\ket{0},
 \end{equation}
 one finds that the evolved density matrix of the probe is given by
 \begin{widetext}
 \begin{equation}\label{reduced}
 	\rho(\tau){=}\dfrac{1}{2}\left(
 	\begin{array}{cc}
 		1+\cos(\theta)e^{-\Gamma_0\tau(2N+1)}-\frac{1-e^{-\Gamma_0\tau(2N+1)}}{2N+1} &
 		\sin(\theta)e^{\frac{-\Gamma_0\tau(2N+1)}{2}-i\tau\Omega+i\phi}\tabularnewline 
 			\sin(\theta)e^{\frac{-\Gamma_0\tau(2N+1)}{2}+i\tau\Omega-i\phi}&
 	1-\cos(\theta)e^{-\Gamma_0\tau(2N+1)}+\frac{1-e^{-\Gamma_0\tau(2N+1)}}{2N+1} 
 	\end{array}
 	\right),
 \end{equation}
\end{widetext}
  where $ N=N(\Omega_0) $.  Moreover,  $\Gamma_0$ {=} $\gamma{(\Omega_0 )}$ denotes the decay coefficient for the atom in the vacuum, $\Omega=\Omega_0 +2\Delta(\Omega_0 )$ represents the Lamb-shifted excitation frequency.

 \section{Quantum thermometry}\label{Quantum Thermometry}
 \par
 We apply the two-level atom for estimating the initial temperature of the quantum field. The information on the temperature is stored into  the evolved state of the total system, and hence is used by the two-level atom to probe it. To this aim, we compute the QFI corresponding to the temperature and analyze its behavior. In particular, we focus on the QFI optimization with respect to initial as well as environmental parameters to achieve optimal thermometry. Using Eqs. (\ref{reduced}) and (\ref{a12}) with $ \lambda=T $, we obtain the following analytical expression for the QFI:

  \begin{widetext}
\begin{align}
H_{T}&=\text{e}^{-2gM}\Bigg(\frac{g \left(4 \cos (\theta ) \left(g M^2 \cos (\theta )+2 g M+2\right)+e^{g M} \left(g M^2 \sin ^2(\theta )-8 \cos (\theta )\right)\right)}{M^2}\nonumber\\
	&-
	\frac{\left(g \sin ^2(\theta )-\frac{2 \left(-g M^2 \cos (\theta )-g M+e^{g M}-1\right) \left(e^{-g M} (M \cos (\theta )+1)-1\right)}{M^3}\right)^2}{\frac{\left(e^{-g M} (M \cos (\theta )+1)-1\right)^2}{M^2}+e^{-g M} \sin ^2(\theta )-1}+\frac{4 \left(g M-e^{g M}+1\right)^2}{M^4}\Bigg)(\dfrac{\partial N}{\partial T})^{2},
\end{align}
 \end{widetext}
where $ g= \gamma{(\Omega_0 )}\tau$ and $ M=2N+1 $.

 \par We focus on two different couplings of the moving system to the thermal bath, and investigate quantum thermometry in two important ranges, i.e., low as well as normal temperatures. 
 The first important result is that the QFI is independent of $ \Omega $, and hence  in both regimes the thermometry is unaffected by the Lamb-shifted excitation frequency. Moreover, we see that the QFI is independent of $\phi  $, and thereby no control over the initial phase is  required to achieve the best estimation of the temperature.
 \subsection{Low-temperature and UDW-coupling regime}
\begin{figure}[ht]
	\subfigure[]{\includegraphics[width=6.1 cm]{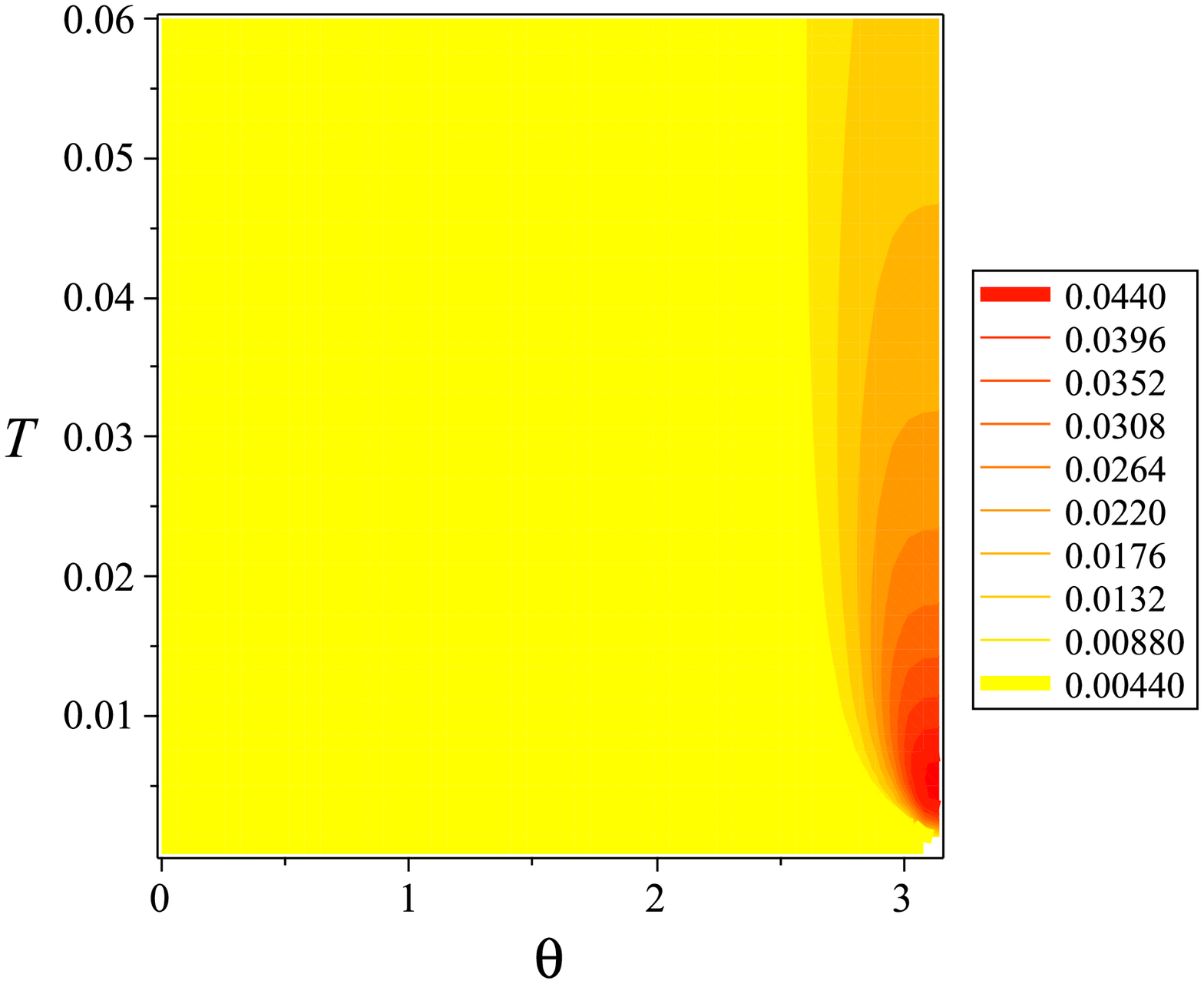}\label{Tvstheta} }
	\hspace{10mm}
	\subfigure[]{\includegraphics[width=6 cm]{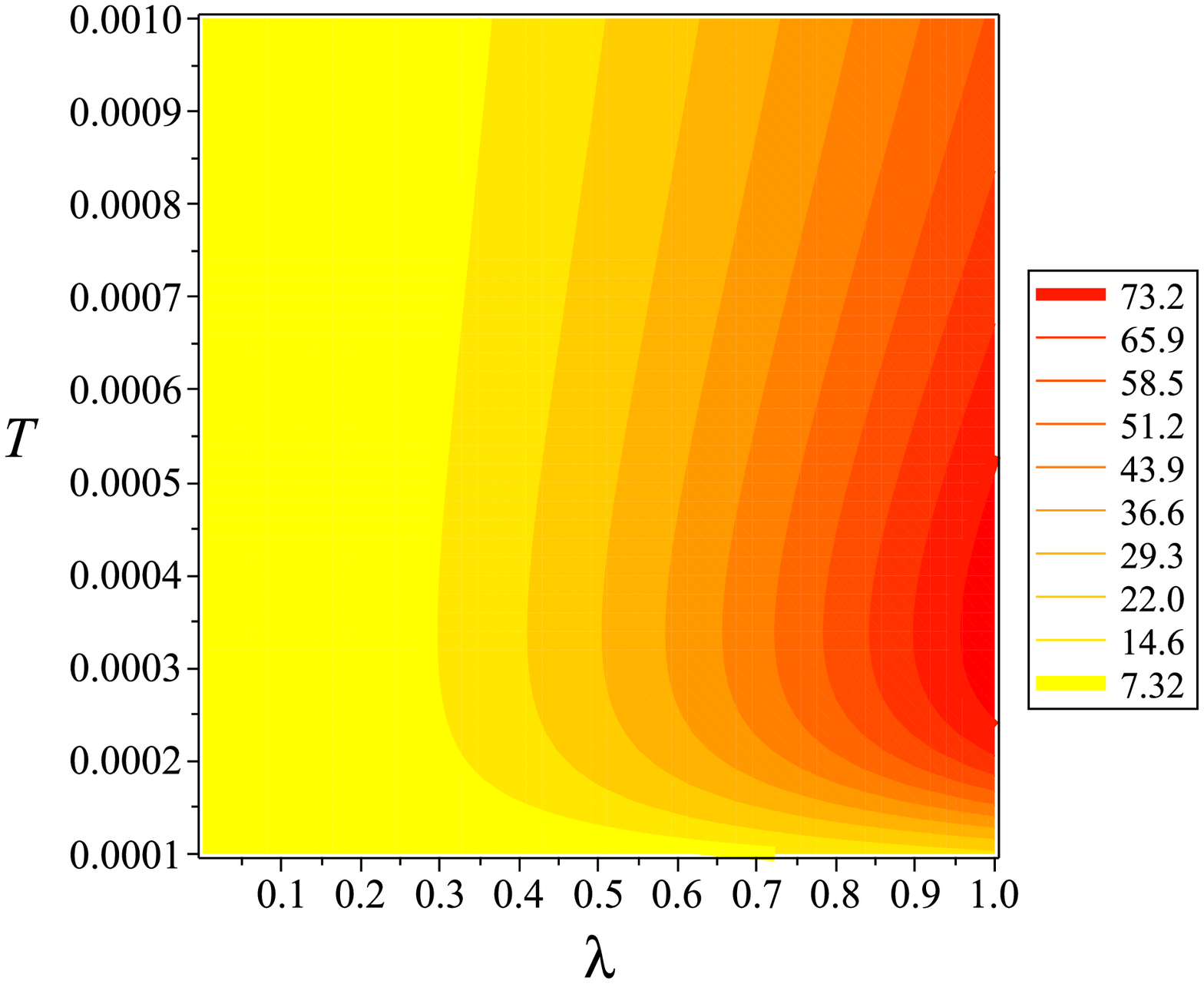}\label{TvsLAMBDA} }
	\caption{UDW-coupling and low-temperature  regime: (a) Quantum Fisher information variation versus temperature T and   weight parameter $ \theta $ for $ u = 4 $, $ \lambda = 0.01$,  and $ \omega = 0.5 $; (b) The same quantity versus $ T $ and coupling constant $ \lambda $  for $ u =5$, and $ \omega=0.1 $.}
	\label{TvsthetaLAMBDA}
\end{figure}

\begin{figure}[hht]
	\subfigure[]{\includegraphics[width=6 cm]{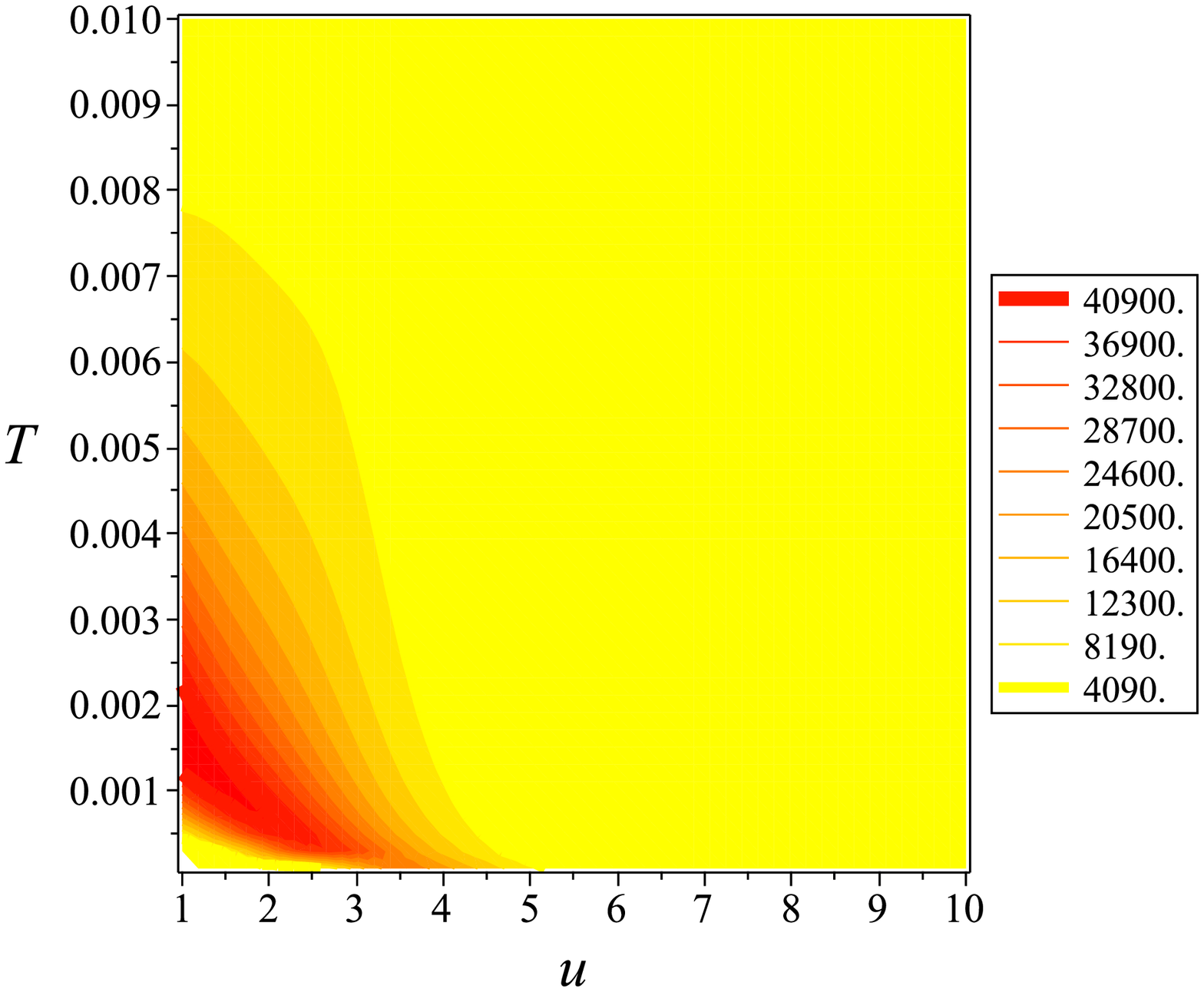}\label{TvsU} }
	\hspace{4.5mm}
	\subfigure[]{\includegraphics[width=5 cm]{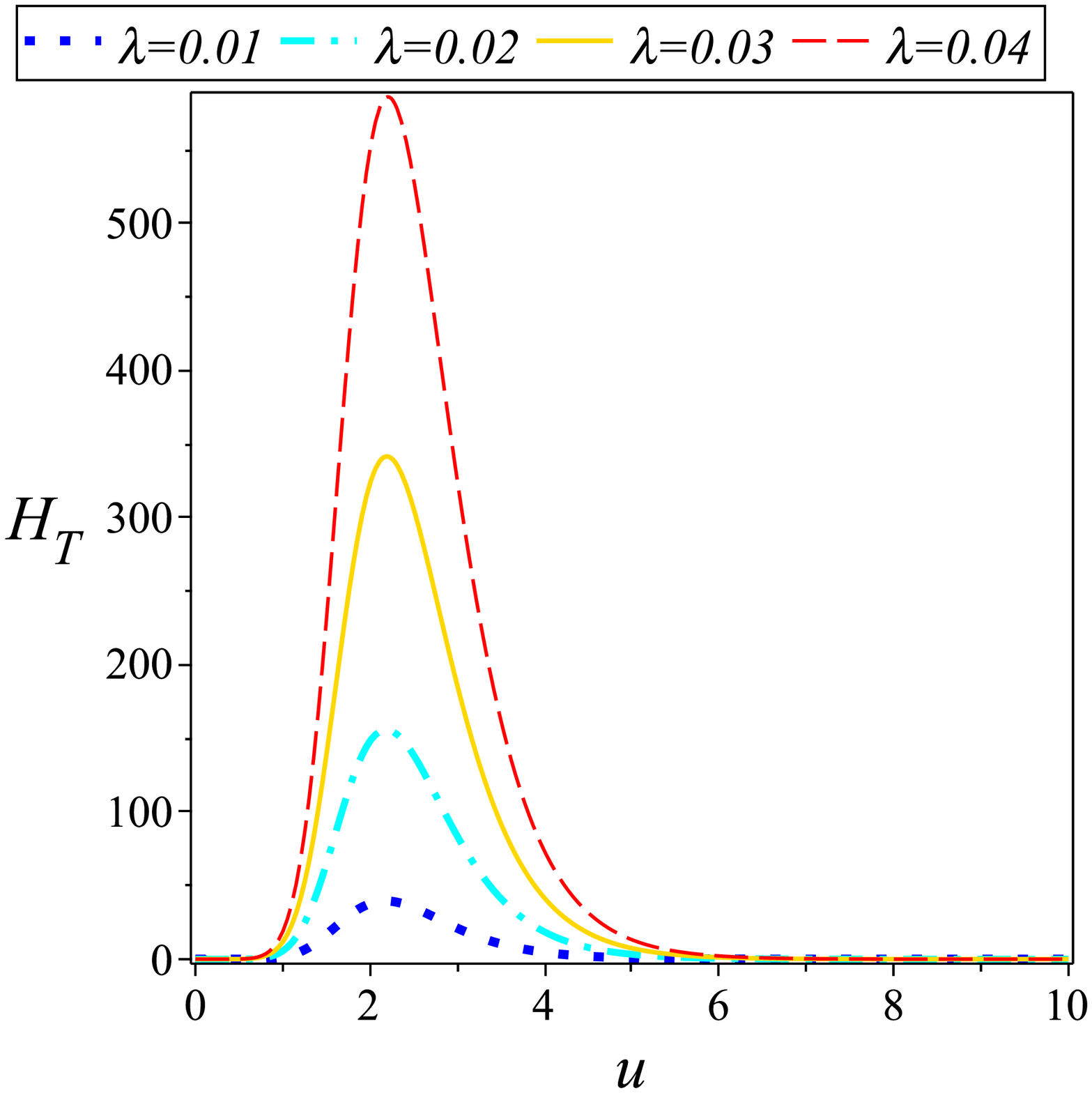}\label{HvsUlAMBDA} }
	\hspace{4.5mm}
	\subfigure[]{\includegraphics[width=5 cm]{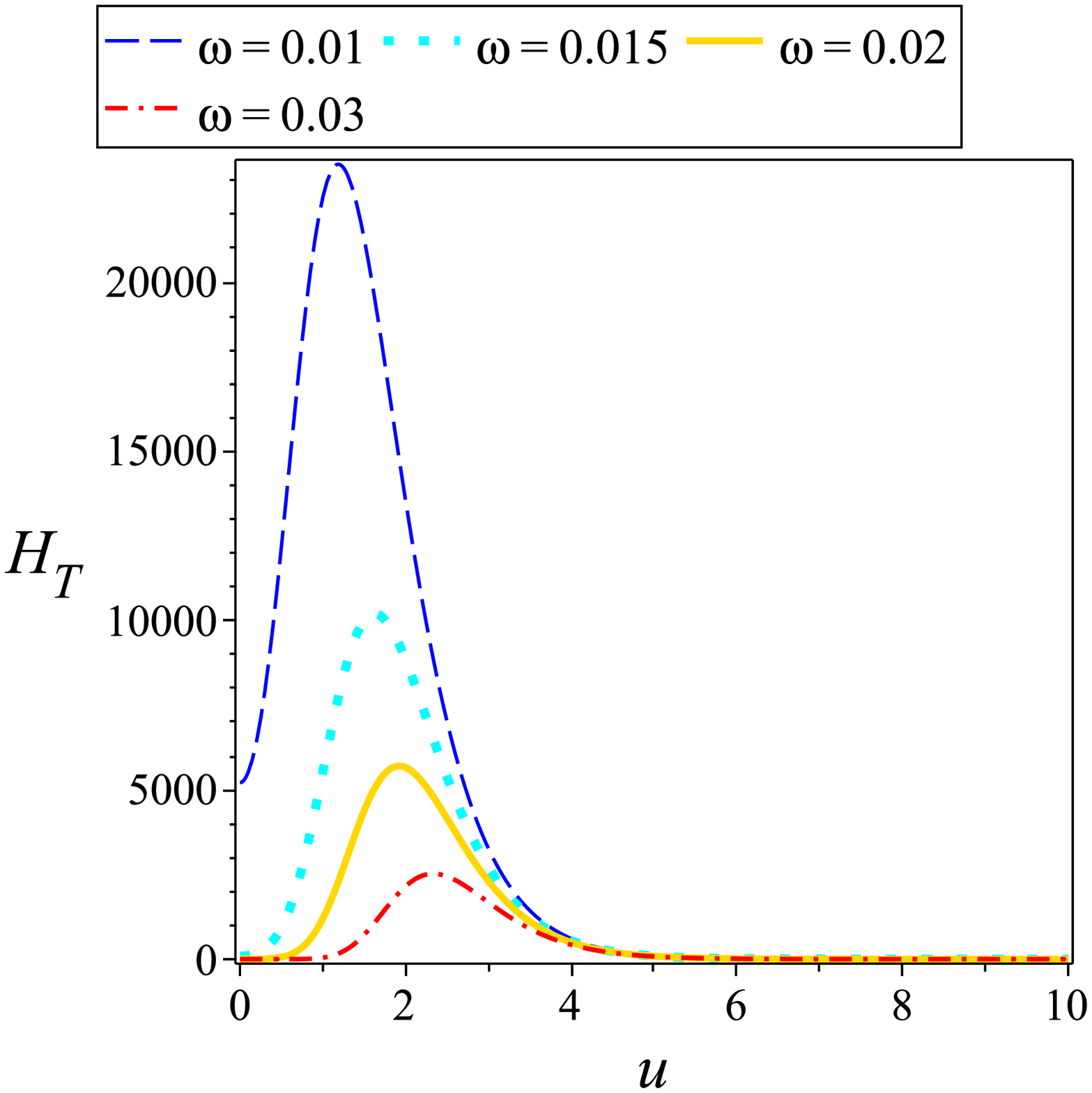}\label{HvsUoMEGA} }
	\caption{UDW-coupling and low-temperature  regime: (a) Quantum Fisher information as a function of  temperature T and  rapidity $ u $ for $ \lambda = 1 $,  and $ \omega = 0.01 $; (b) Quantum Fisher information variation versus $ u $ for $ T=0.001 $,  $ t =100000 $, $ \omega= 0.03 $, and different values of $ \lambda $; (c) The same quantity as a function of  rapidity $ u $ for  $ T= 0.001 $,  $ \lambda = 0.1 $ and different values of $ \omega $.}
	\label{HvsUOL}
\end{figure}

\begin{figure}[hht]
	\subfigure[]{\includegraphics[width=5 cm]{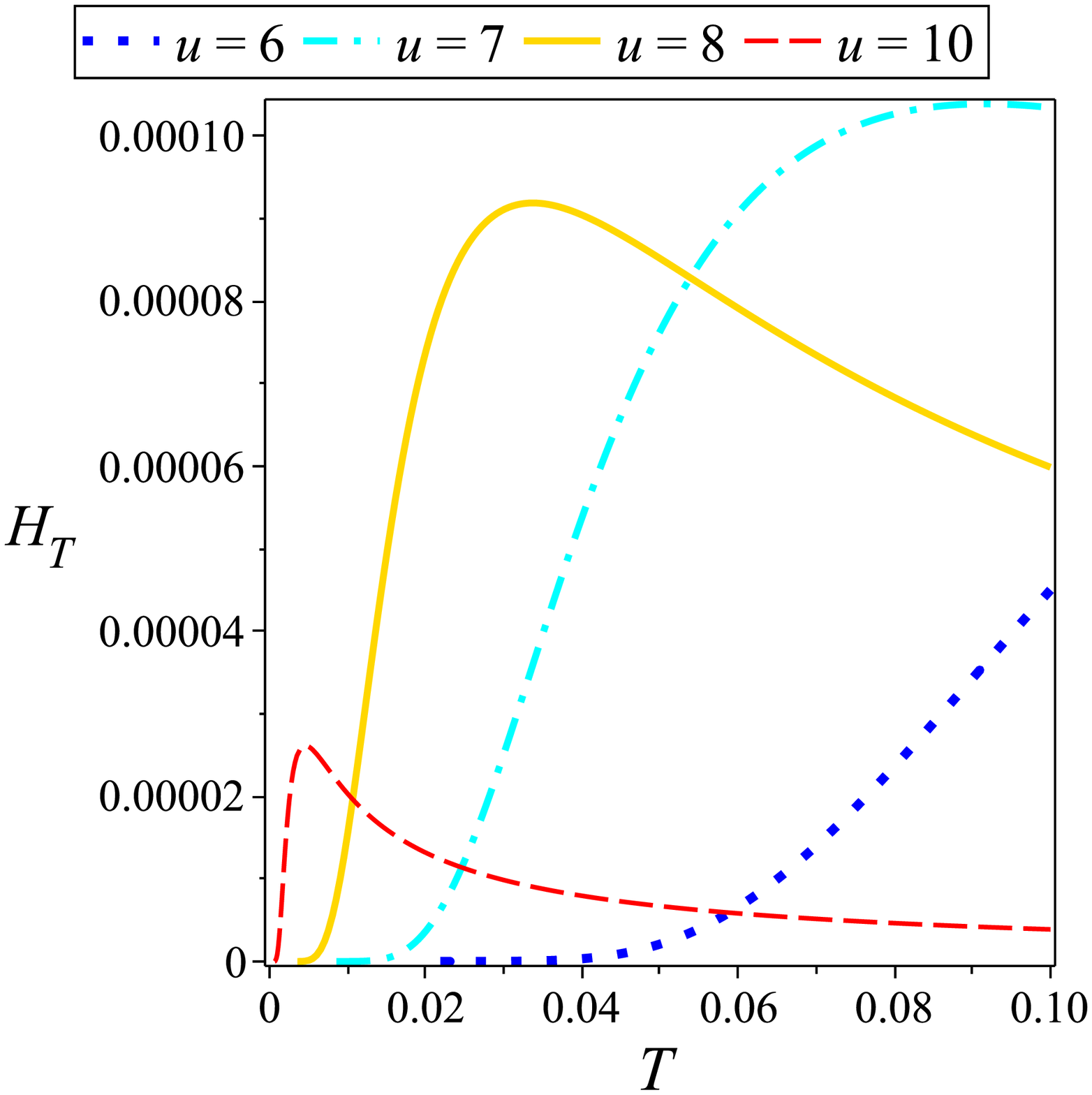}\label{HvsTforU} }
	\hspace{4.5mm}
	\subfigure[]{\includegraphics[width=6 cm]{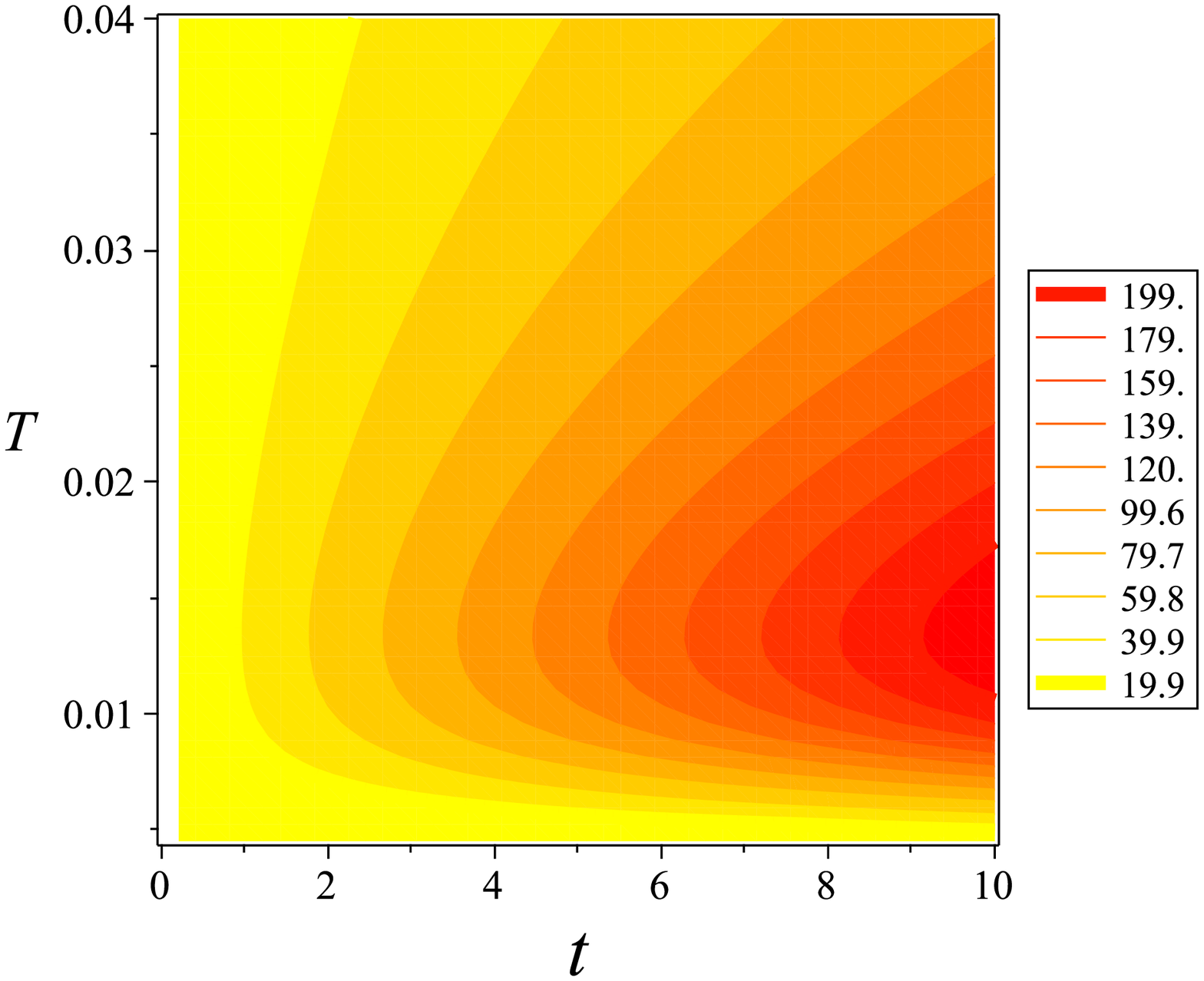}\label{HvsTContour} }
	\hspace{4.5mm}
	\subfigure[]{\includegraphics[width=5 cm]{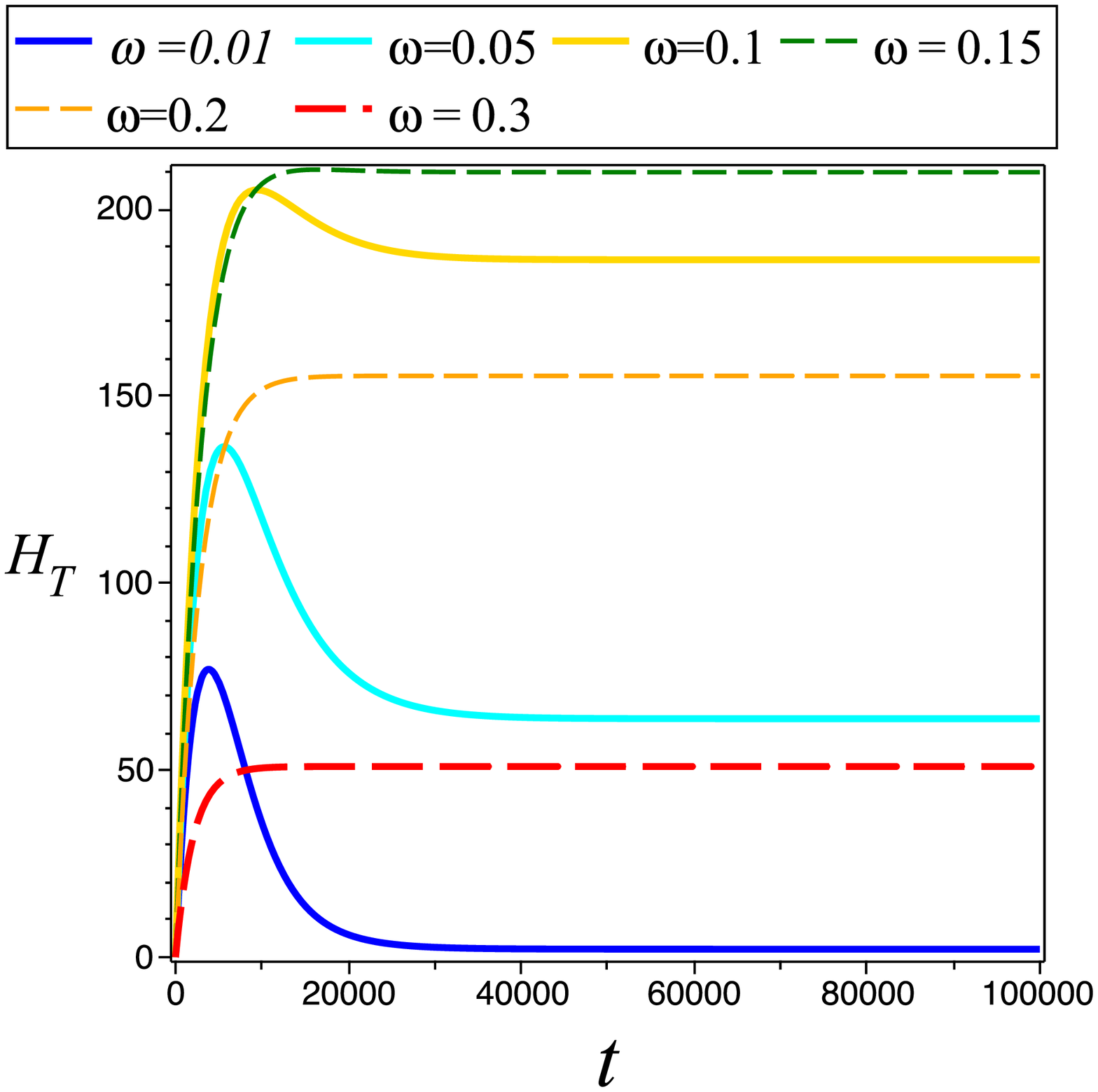}\label{HvstforOmega} }
	\caption{UDW-coupling and low-temperature  regime: (a) Quantum Fisher information as a function of   temperature T  for $ \lambda = 0.1 $,  $ \omega = 200 $  and  different values of rapidity $ u $ ; (b) The same quantity versus  $ T $ and $ t $ for  $ u=0.01 $, $ \lambda=1 $,  and $ \omega= 0.05$; (c) The same quantity versus  $ t$ for  $ T= 0.05 $, $  \lambda= 0.1$,  $ u= 0.1 $ and different values of $ \omega $.}
	\label{HvsTtforUContourforOmega}
\end{figure}

\begin{figure}[hht]
	\subfigure[]{\includegraphics[width=5 cm]{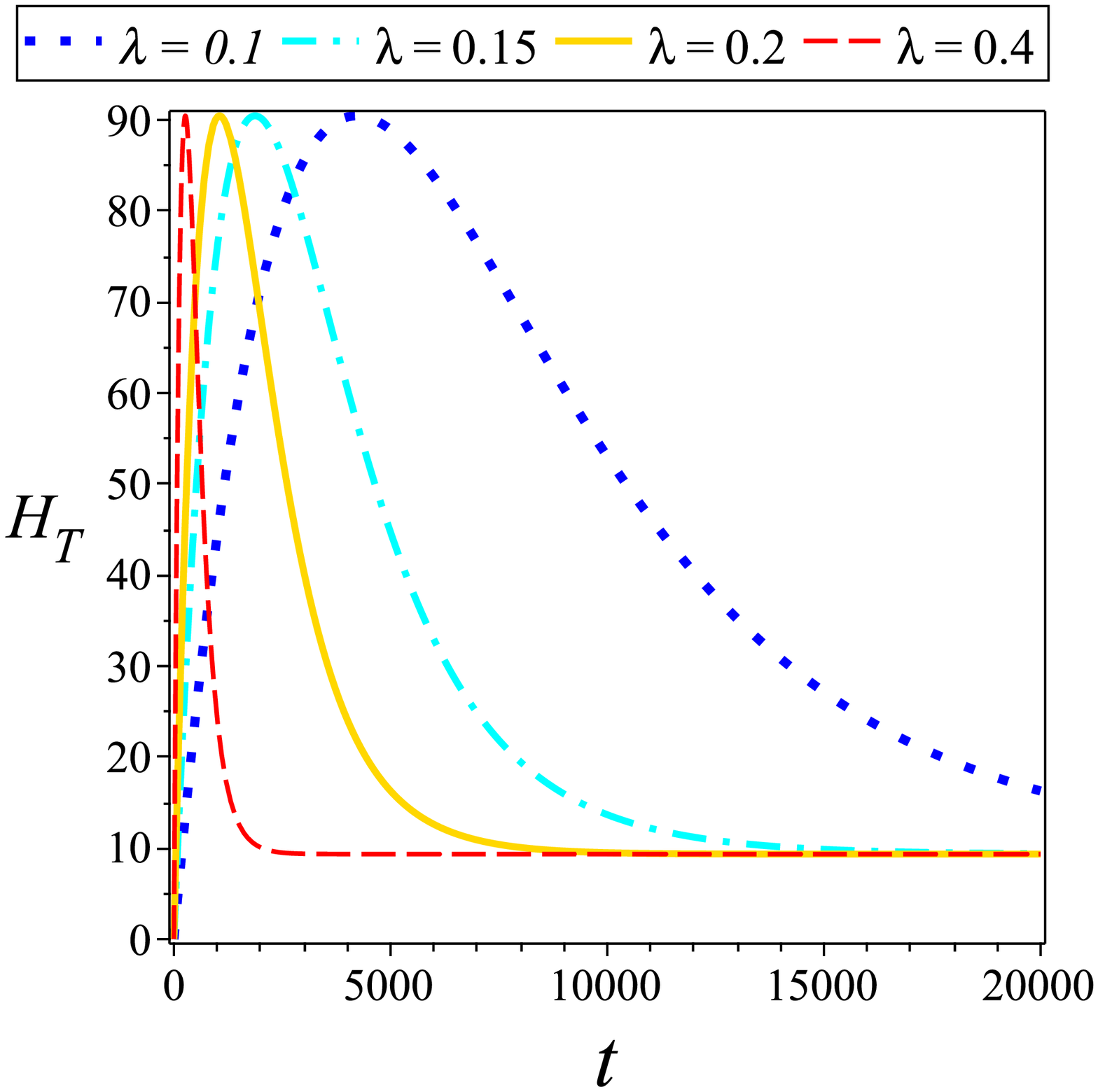}\label{Hvstdiflambdaweak} }
	\hspace{4.5mm}
	\subfigure[]{\includegraphics[width=5cm]{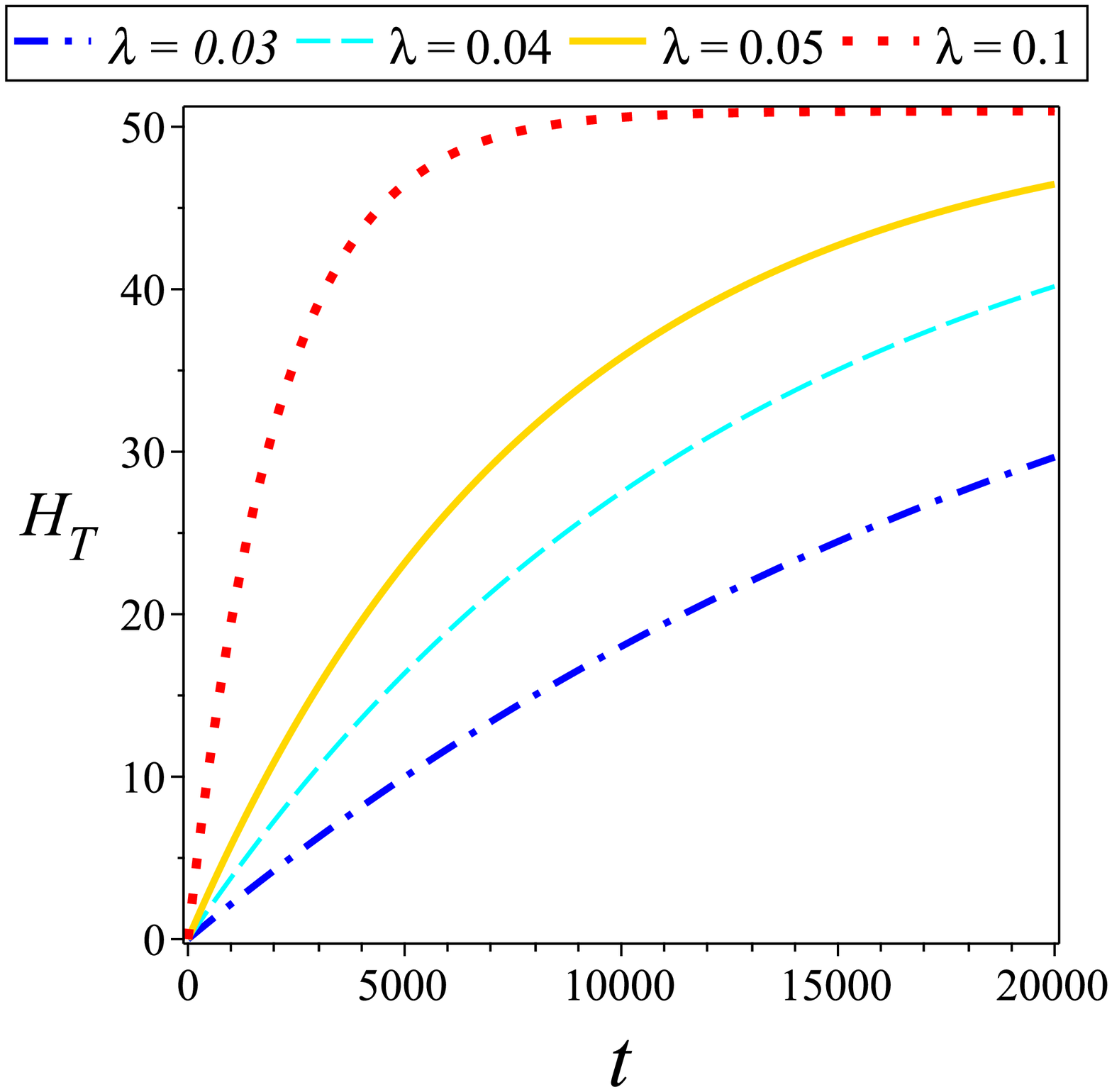}\label{Hvstdiflambdastrong} }
	\hspace{4.5mm}
	\subfigure[]{\includegraphics[width=5 cm]{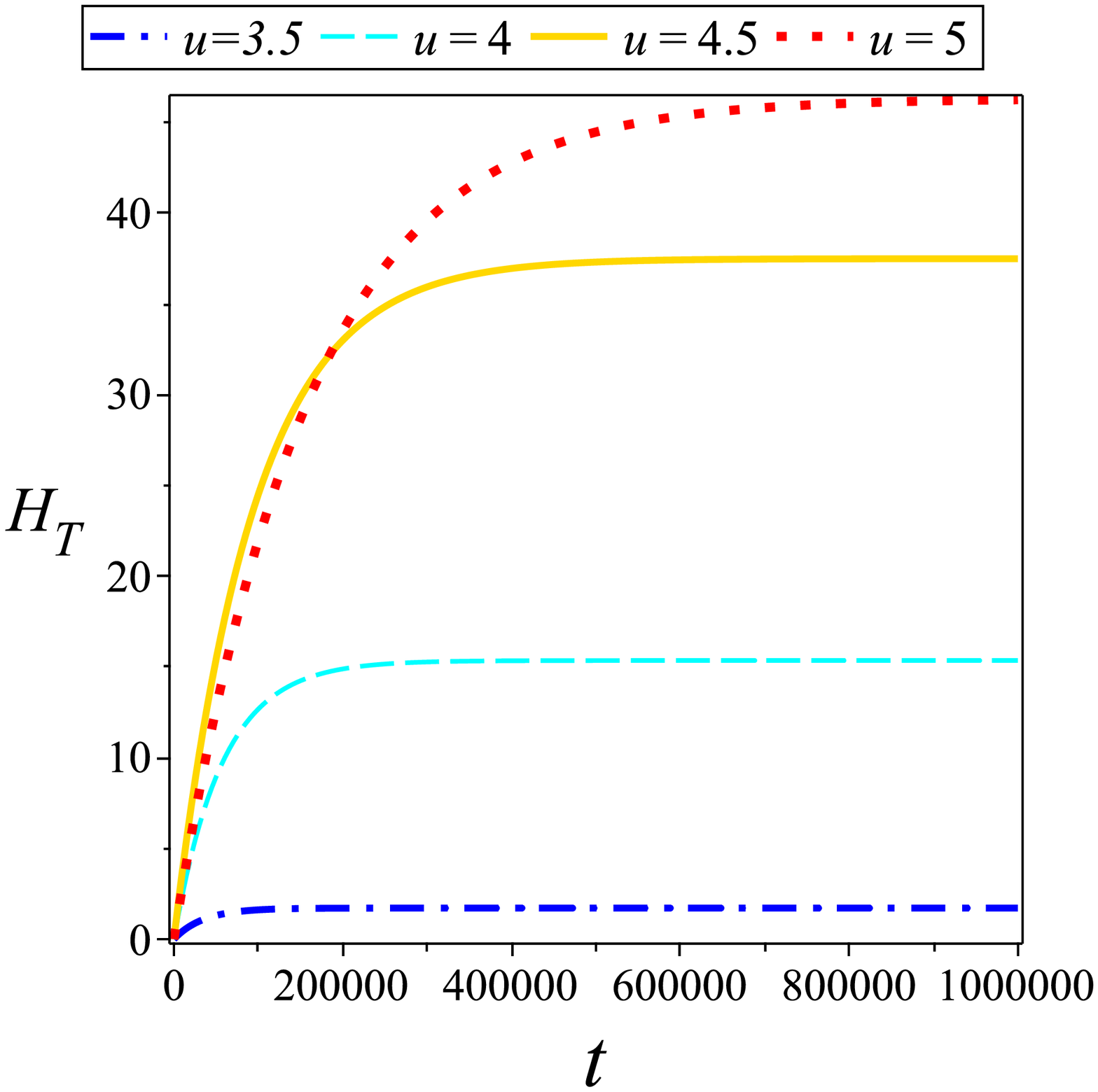}\label{Hvstdifustrong} }
	\caption{UDW-coupling and low-temperature  regime: (a) Dynamics of the quantum Fisher information (QFI) with respect to the temperature   in the  absence  of  QFI trapping for $ T=0.05 $, $ \omega=0.02, u=0.1 $, and different values of $ \lambda $;  (b) The same quantity, plotted in the presence of the QFI trapping  for $ T=0.05 $, $ \omega=0.3, u=0.1 $;  (c) The QFI dynamics   in the presence of the QFI  trapping for $ T=0.001 $, $ \omega=0.3, \lambda=0.1 $, and different values of  $ u $.}
	\label{Hvstdiflambdaweakstrong}
\end{figure}
 Studying the QFI behavior for $ 0<T \ll 1 $ and UDW coupling, we find that when the atom is initially prepared in the ground state $ (\theta=\pi) $, the best estimation can be achieved (see Fig. \ref{Tvstheta}). Therefore, we set $ (\theta=\pi) $ throughout this paper.  Moreover, as demonstrated in Fig. (\ref{TvsLAMBDA}), when the coupling constant $ \lambda $, quantifying the strength of interaction between the probe and the field, increases, the QFI rises, enhancing the accuracy of the quantum estimation.
 
 \par
 Figure \ref{TvsU} illustrates how the rapidity of the trajectory affects the thermometry accuracy at different low temperatures. It shows that the QFI first improves with an increase in $ u $ and then decreases. Although strengthening the coupling constant, one can enhance the estimation, we see from Fig. \ref{HvsUlAMBDA} that,  in the  weak-coupling  regime,  an increase in $ \lambda $, cannot shift the optimal value of $ u $ at which the best estimation occurs. 
 In addition, Fig. \ref{HvsUoMEGA} shows that an increase in $  \omega $ reduces the accuracy of the optimal estimation and shifts the optimal value of $ u $, at which the best estimation occurs, to higher velocities.
 
 \par
 As clear in Figs. \ref{TvsthetaLAMBDA} and \ref{TvsU}, the QFI first grows with an increase in the temperature and then falls. Figure \ref{HvsTforU} illustrates how the variation of the probe velocity affects this fall. It shows that by increasing $ u $,  the optimal value of the QFI versus T is achieved in lower temperatures. However, it may cause the optimal value to decrease, thus reducing the precision of the estimation. 
 
Monitoring the dynamics of the quantum thermometry through studying the QFI shows when time goes on, the QFI improves, as expected, however,  then it decays (see Fig. \ref{HvsTContour} ). Control of this reduction, studied in Figs.  \ref{HvstforOmega} and \ref{Hvstdiflambdaweakstrong}, is of great interest.  
Figure  \ref{HvstforOmega} displays that an increase in $ \omega $ removes the QFI dropping with time and results in the QFI trapping exhibiting asymptotic behavior with some definite value.  However, after the appearance of the QFI trapping, a further increase in $ \omega $ suppresses the QFI, although it causes the QFI trapping to appear sooner.
Therefore, to investigate the QFI dynamics in the low-temperature scenario, the low-frequency regime, where the QFI exhibits an optimum point,  and the high-frequency one, in which the QFI trapping occurs, should be studied separately.

Figure \ref{Hvstdiflambdaweak}  illustrates a positive and interesting effect of a rise in the coupling constant. It demonstrates that when it is raised in the low-frequency regime, the optimal value of the QFI, signifying the best instance for quantum thermometry, is achieved sooner. Nevertheless, this strategy brings the QFI loss forward, and hence the period at which the thermometry can be implemented efficiently is shortened.

The QFI variations versus time for different values of $ \lambda $ and $ u $ in the high-frequency regime are displayed in Figs. \ref{Hvstdiflambdastrong} and \ref{Hvstdifustrong}, respectively. Clearly, when the interaction between the probe and field is strengthened by an increase in $ \lambda $,  the estimation enhances and it causes the QFI trapping  to appear more quickly. Moreover, as clear from Fig. \ref{Hvstdifustrong}, speeding up the probe, applied for the thermometry, retards the QFI.

\subsection{Low-temperature and TD-coupling regime}
\begin{figure}[hht]
		\subfigure[]{\includegraphics[width=5 cm]{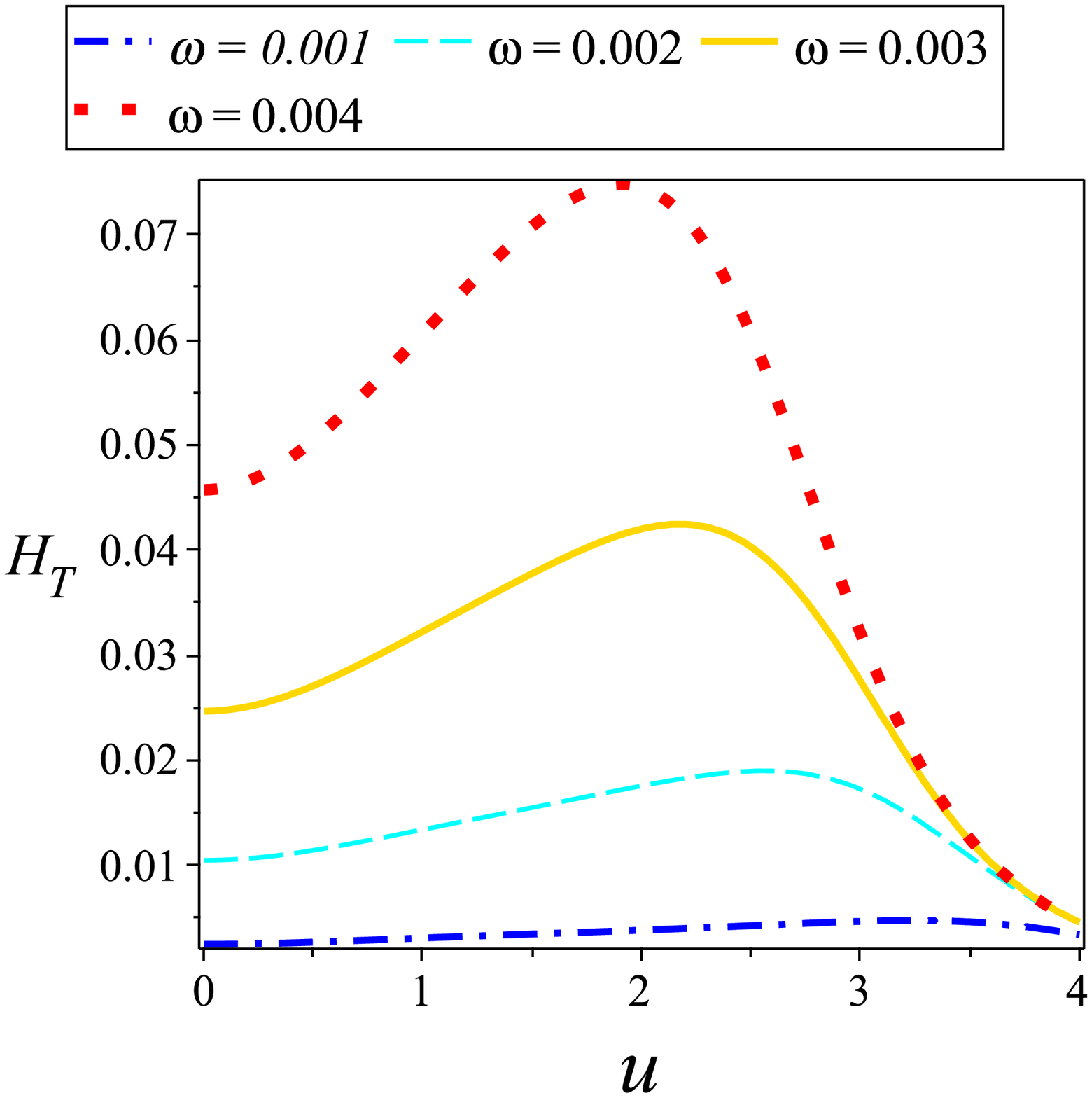}\label{HTvsUdiffOTD} }
		\hspace{4.5mm}
	\subfigure[]{\includegraphics[width=5 cm]{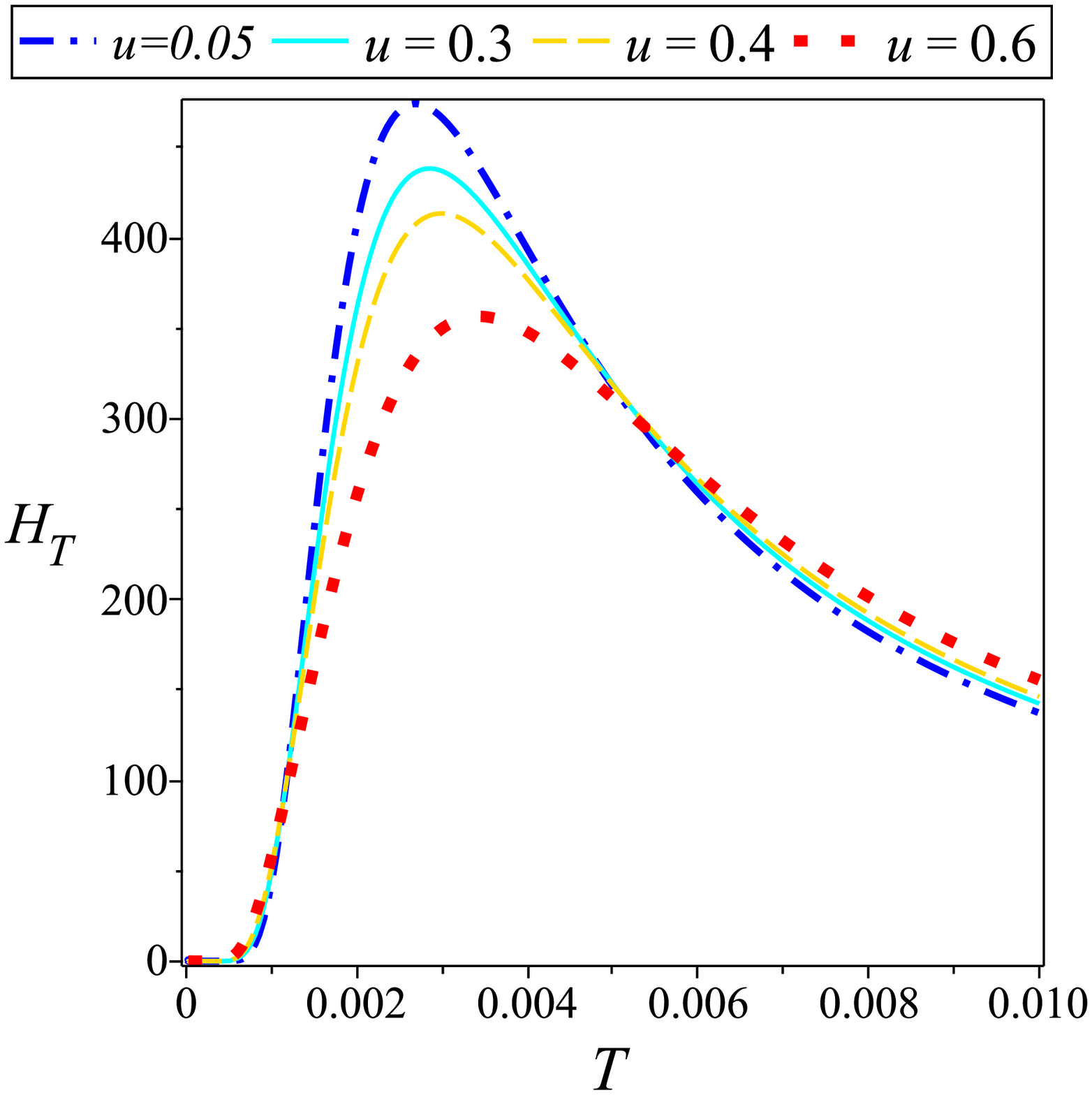}\label{HTvsTdiffUtD} }
	\hspace{4.5mm}
	\subfigure[]{\includegraphics[width=5cm]{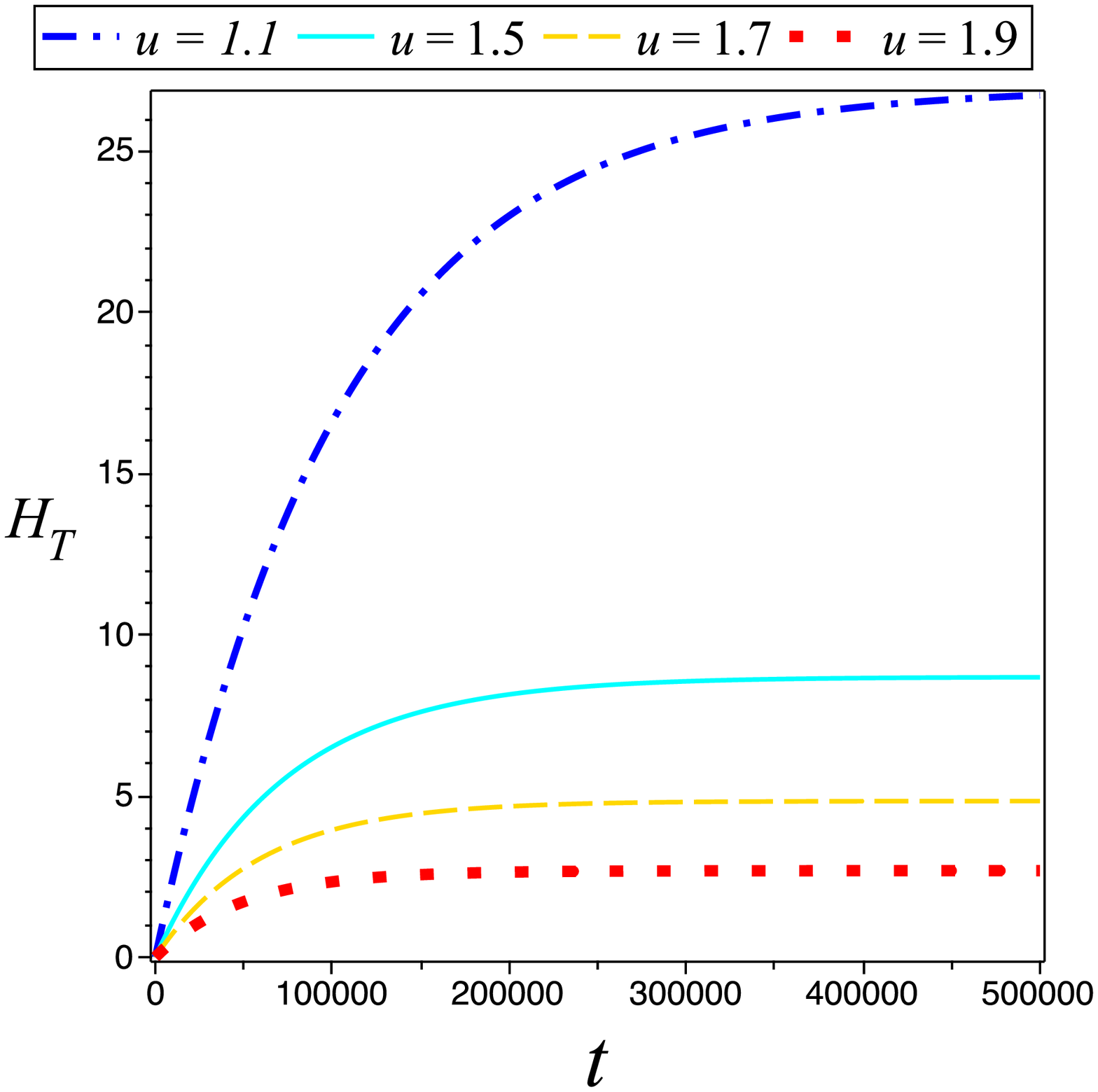}\label{HTvstdiffUTDD} }

	\caption{TD-coupling and low-temperature  regime: (a) Quantum Fisher information  as a function of  rapidity $ u $ for  $ T= 0.01 $,  $ \lambda = 0.01 $ and different values of $ \omega $; (b) The same quantity versus $ T $ for $ \omega= 0.01 $,  $ \lambda = 0.2 $, and different values of $ u $; (c) The QFI dynamics in the presence of the QFI trapping for $T=0.05,  \omega= 0.2 $,  $ \lambda = 0.06 $ , and different values of $ u $.}
	\label{HTvsTTD}
\end{figure}
In this section, we focus on the quantum thermometry when the temperature is sufficiently low in the TD-coupling regime. Most of the results presented for the UDW coupling also hold here, and hence we only present the deviations.
\par
In the TD-coupling regime, although the best estimation is achieved for $ \theta=\pi $, we see that at high velocities, the QFI does not vary considerably with $ \theta $. Therefore, when the probe moves at high speeds, it is not necessary to initially prepare it in the ground state to achieve the best estimation.

 Moreover, in the UDW-coupling we saw that enhancement of the thermometry was limited to certain circumstances. For example, strengthening the interaction between the probe and the field does not necessarily lead to the estimation improvement in the low-frequency and UDW -coupling regime. However, in the TD coupling, we find that an increase in $ \lambda $  always improves the accuracy of the temperature estimation.
 
 \par
 Figure \ref{HTvsUdiffOTD} illustrates that an increase in $ \omega $ raises the accuracy of the optimal thermometry and shifts the optimal value of $ u $, at which the best estimation is achieved, to lower velocities.
 
 \par
 Another difference between the UDW- and TD-coupling regimes is demonstrated in Fig. \ref{HTvsTdiffUtD} exhibiting how the variation of the probe velocity influences the QFI decay occurring with an increase in the temperature.   We see that speeding up the probe negatively affects low-temperature thermometry, because it shifts the optimal value of the QFI to higher temperatures. Similar to UDW coupling, we find that raising  $ u $ suppresses the optimal value of the QFI versus $ T $,  reflecting that the thermometry becomes more inaccurate. Moreover, investigating the QFI dynamics at high frequencies in which the QFI trapping occurs reveals that a decrease in $ u $ retards the QFI trapping (see Fig. \ref{HTvstdiffUTDD}).

  \subsection{Normal-temperature and UDW-coupling regime}
  \begin{figure}[hht]
  	\subfigure[]{\includegraphics[width=5 cm]{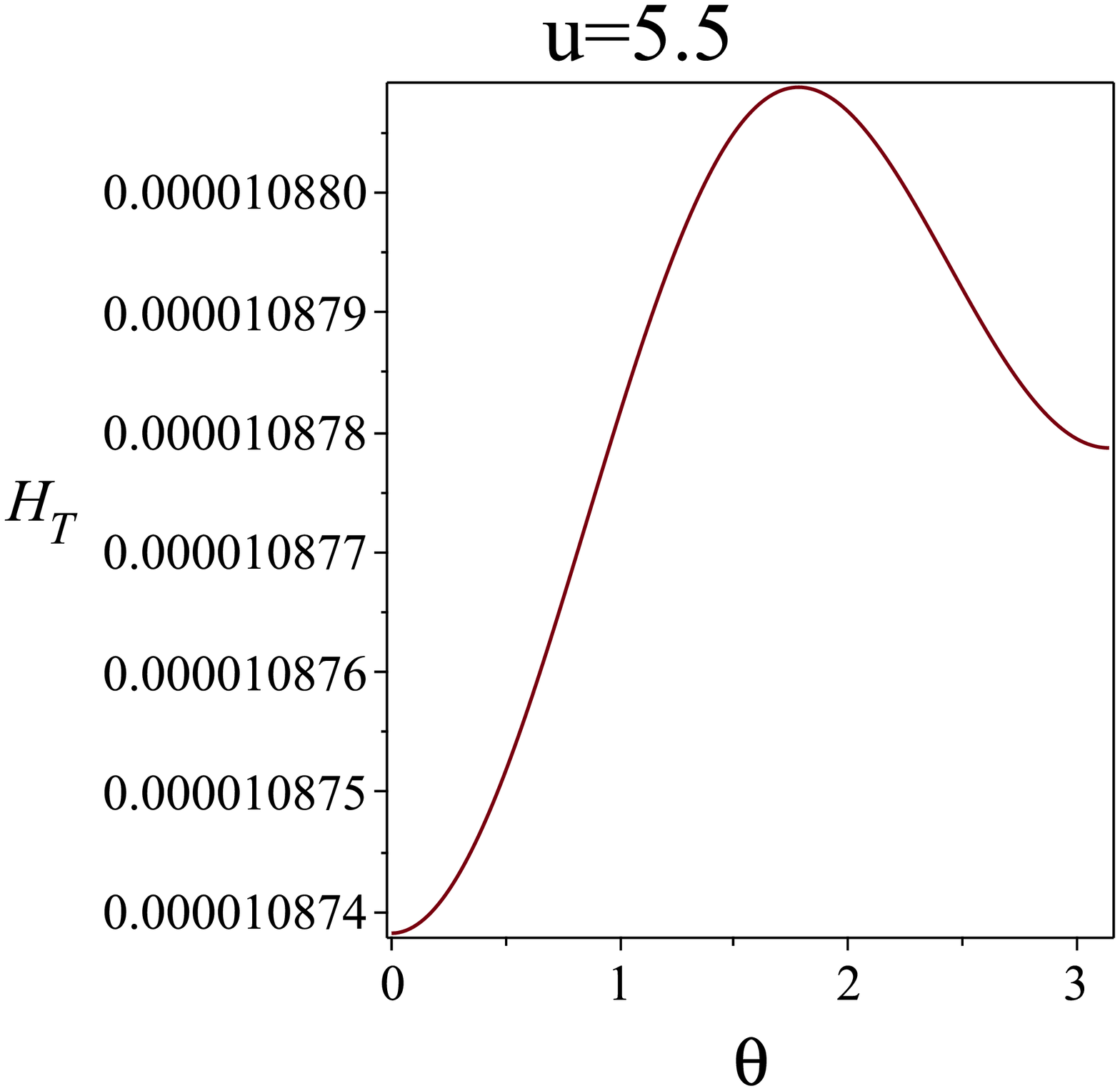}\label{NormalHthetau1} }
  	\hspace{4.5mm}
  	\subfigure[]{\includegraphics[width=5 cm]{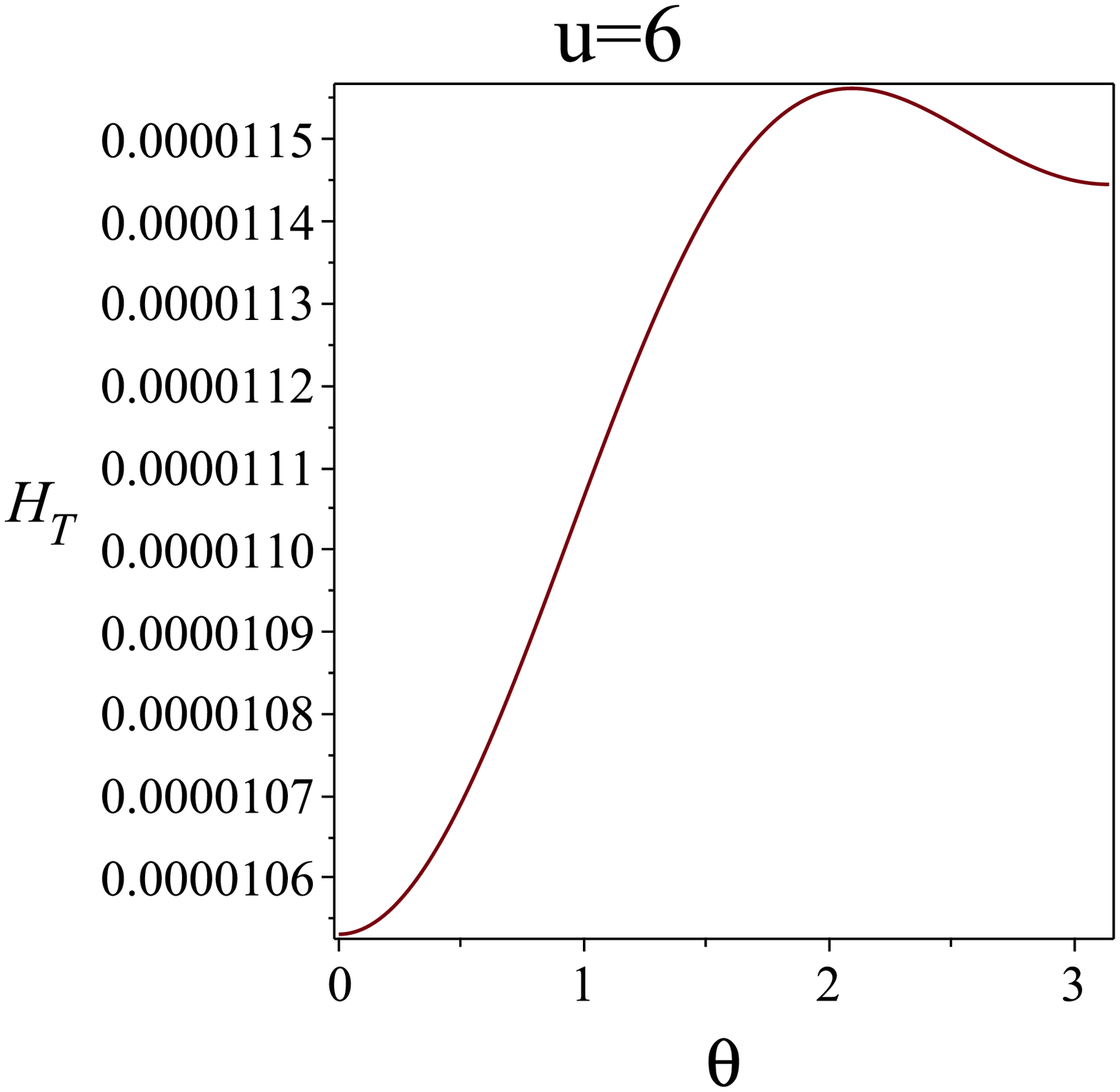}\label{NormalHthetau2} }
  	\hspace{4.5mm}
  	\subfigure[]{\includegraphics[width=5cm]{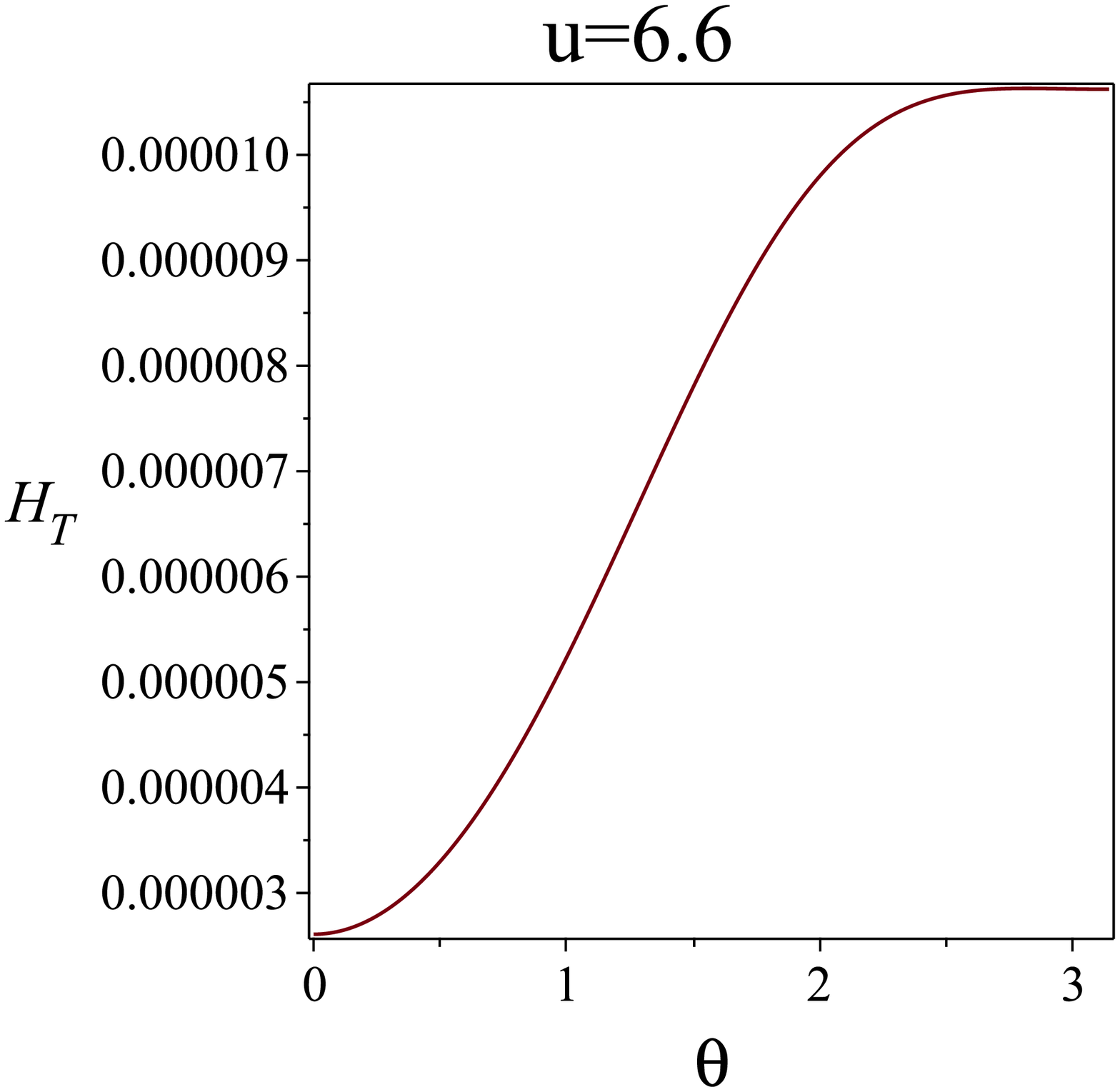}\label{NormalHthetau3} }
  	\caption{UDW-coupling and normal-temperature  regime: quantum Fisher information  versus  $ \theta$ for  $ T= 100 $,  $ \lambda = 0.1,  \omega =10$ and different values of $ u $.}
  	\label{NormalHthetau}
  \end{figure}

\begin{figure}[ht]
	\subfigure[]{\includegraphics[width=8.5cm]{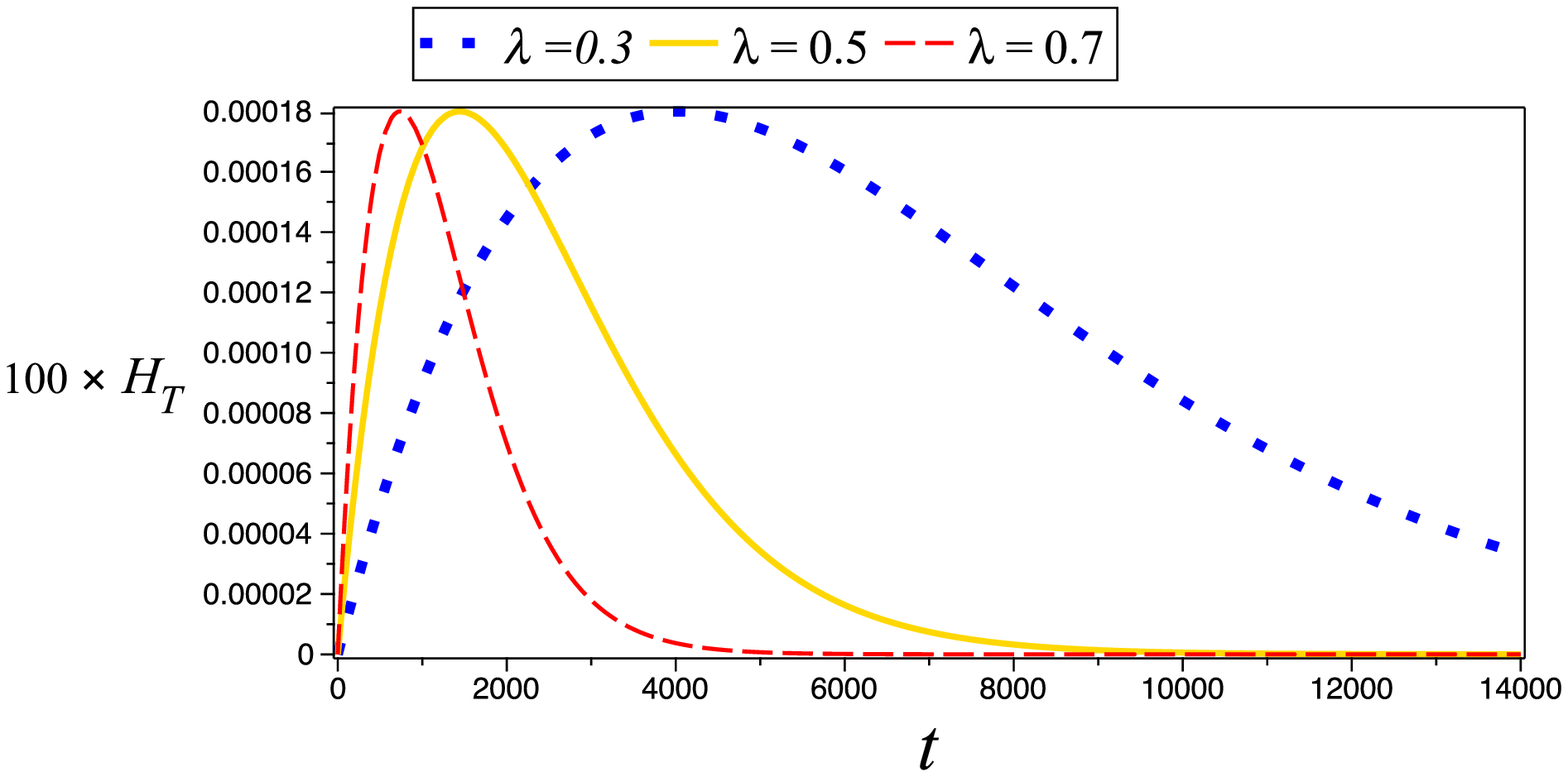}\label{NormalHtLambda} }
	\hspace{4.5mm}
	\subfigure[]{\includegraphics[width=8.5cm]{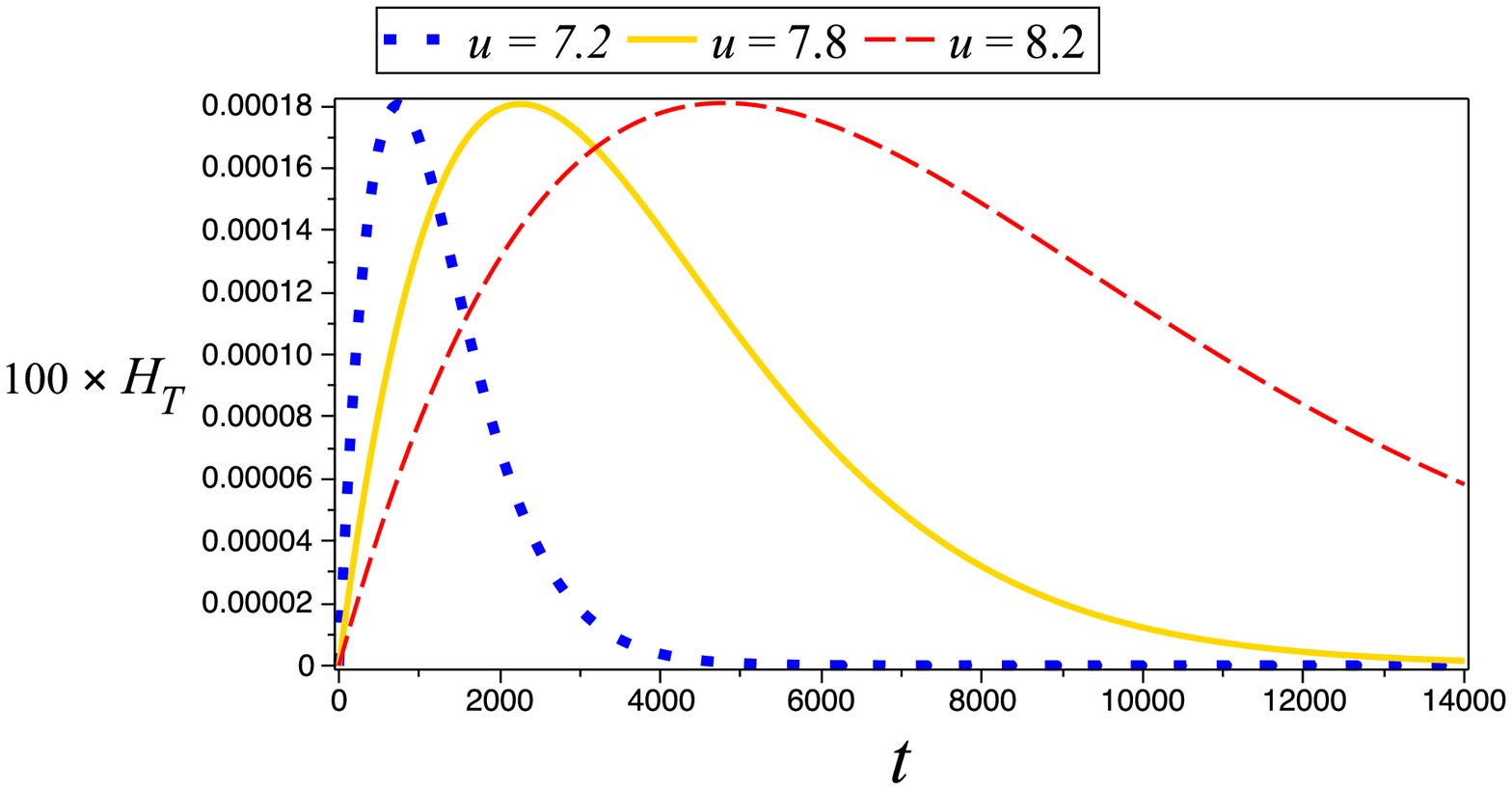}\label{NormalHtU} }

	\caption{UDW-coupling and normal-temperature  regime: (a) Dynamics of the quantum Fisher information   for  $ T= 300 $, $ u=7.2, \omega =0.01$ and different values of $ \lambda $. (b) The same quantity for $ T= 300 $, $ \lambda=0.7, \omega =0.01$ and different values of $ u $. }
	\label{NormalHtLambdaU}
\end{figure}

\begin{figure}[ht]
	\subfigure[]{\includegraphics[width=8.5cm]{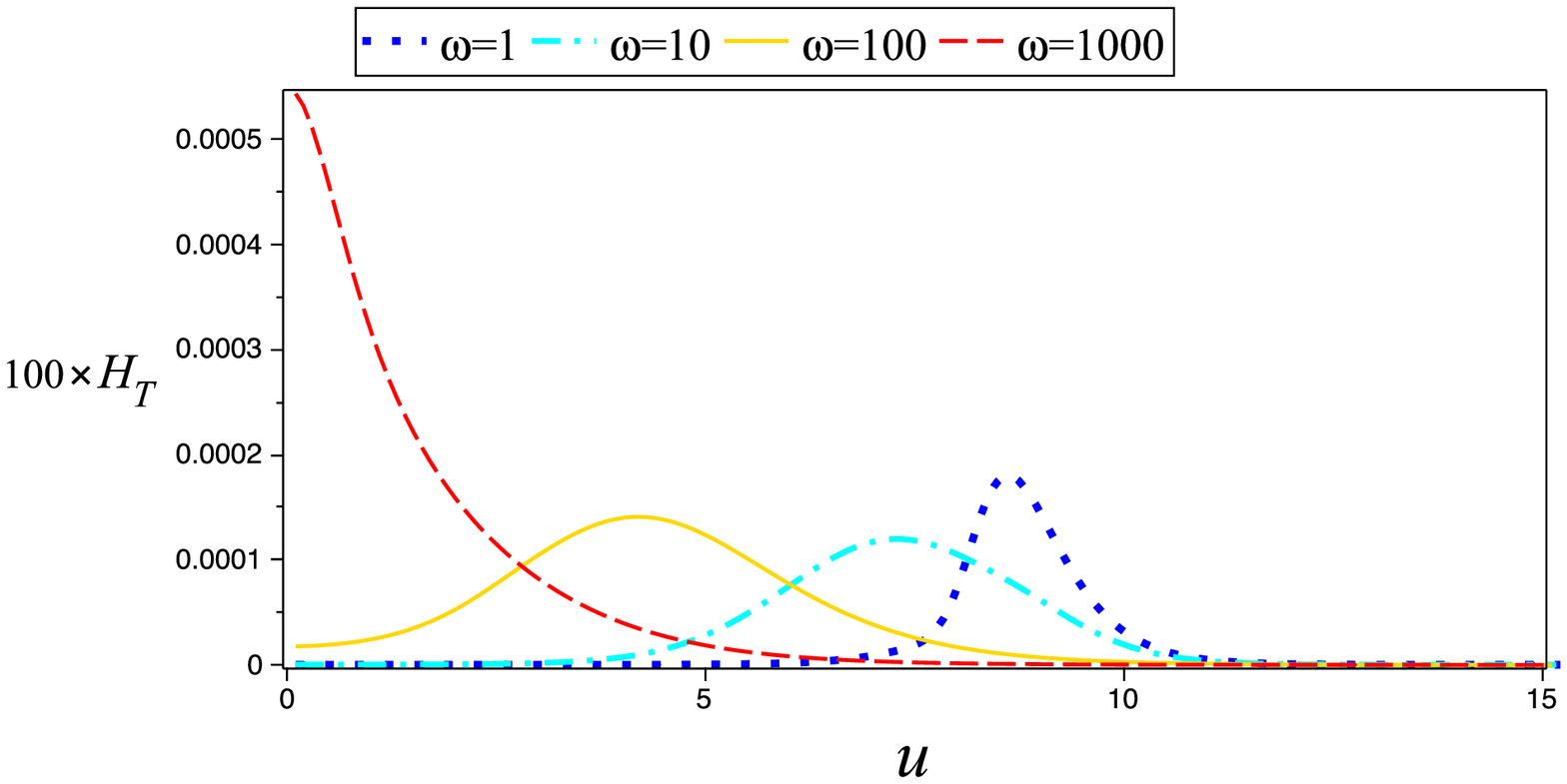}\label{NormalHuOmega1} }
	\hspace{4.5mm}
	\subfigure[]{\includegraphics[width=8.5cm]{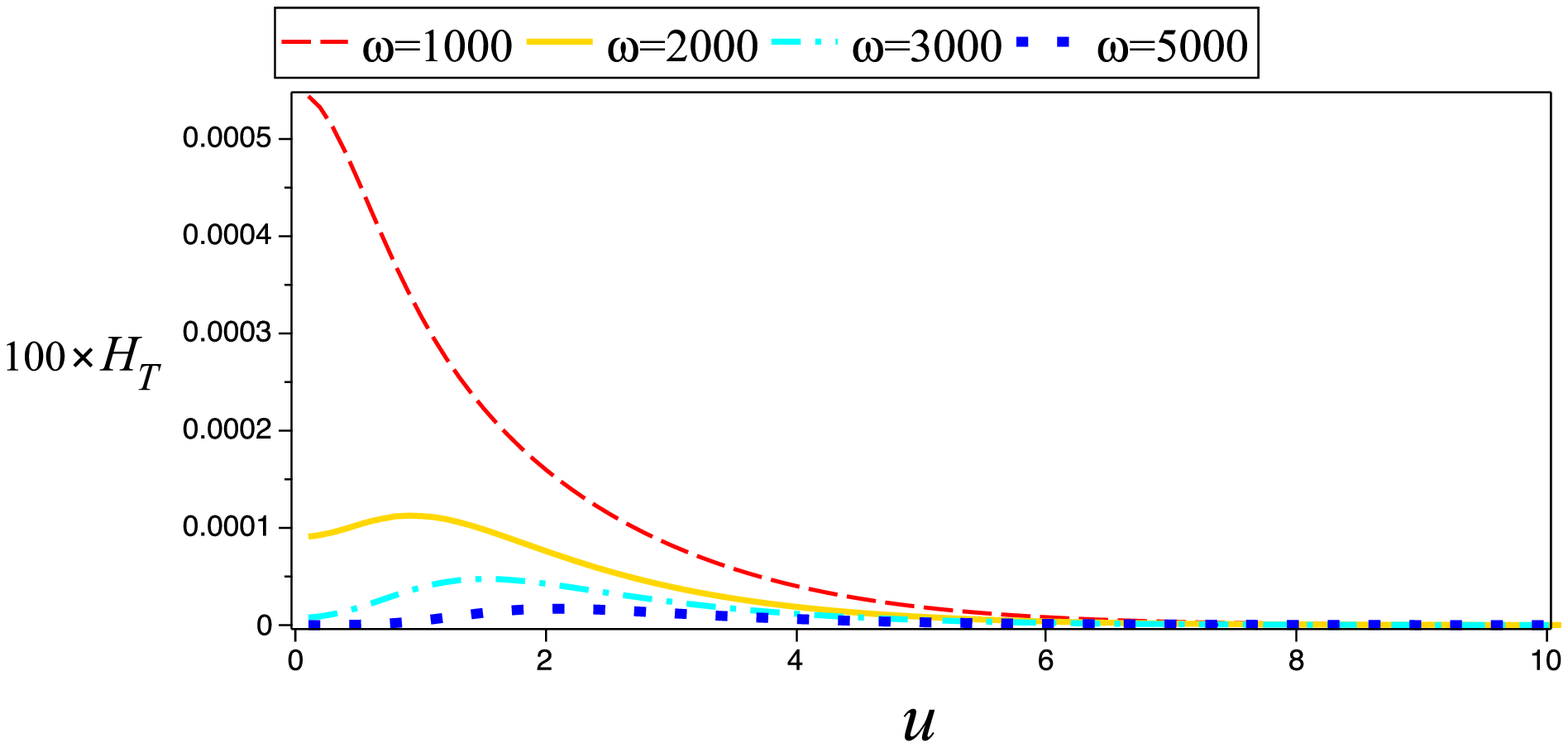}\label{NormalHuOmega2} }
	
	\caption{UDW-coupling and normal-temperature  regime: quantum Fisher information vesus $ u $  for  $ T= 300 $, $ \lambda=2.5$ and different values of (a) low and (b) high frequencies.}
	\label{NormalHuOmega12}
\end{figure}
Now we investigate the normal-temperature thermometry in the range $ 1<T\sim 300 $ and present the most important results. First, we find that, at high velocities, the best estimation is achieved for $ \theta=\pi$ (see Fig. \ref{NormalHthetau}). Therefore, probes moving at high speeds should be initially prepared in the ground state to implement the optimal thermometry. Moreover, similar to the low-temperature regime, to study the QFI dynamics, we should consider two different scenarios: 1) the low-frequency regime in which the QFI first increases with time and then decreases; 2) the high-frequency regime where the QFI trapping occurs.

\par
For high frequencies,  we find that the QFI dynamics is similar to one observed in the low-temperature regime. Therefore, 1) an increase in $\omega$ and $\lambda$ causes the QFI trapping to appear sooner; 2) a rise in $\lambda$ improves the QFI; 3) an increase in u retards the QFI trapping.

\par
For low frequencies in which the QFI is suppressed with time, an increase (a decrease) in $ u~(\lambda) $   retards the QFI loss during the evolution and hence enhances the estimation of the parameter at periods in which the QFI tends to zero. However, a decrease (an increase) in $ u~(\lambda) $  leads to the occurrence of the optimal estimation at an earlier time. In particular, it does not vary the optimal value of the QFI. These results are exhibited in Figs. \ref{NormalHtLambda}
and \ref{NormalHtU}.

\par
In the low-frequency regime, investigating the QFI behavior versus $u$, we find that an increase in $\omega$ leads to the appearance of the optimal estimation for lower velocities (see Fig. \ref{NormalHuOmega1} ). However, as demonstrated in Fig. \ref{NormalHuOmega2}, in the high-frequency regime, an increase in $\omega$ suppresses the QFI.

 \subsection{Normal-temperature and TD-coupling regime}
 \begin{figure}[ht]
 	\includegraphics[width=8.7cm]{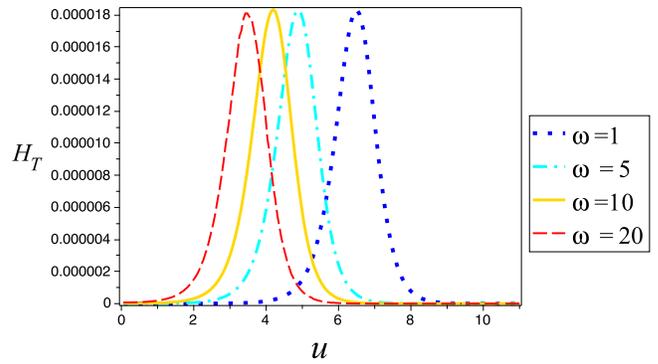}

 	\caption{TD-coupling and normal-temperature  regime: quantum Fisher information vesus $ u $  for  $ T= 160 $, $ \lambda=0.06$ and different values of $ \omega $.}\label{NormalHTuOmega} 
 
 \end{figure}
 Focusing on TD-coupling and normal-temperature regime, we see that the results extracted from Figs.  \ref{NormalHthetau} and \ref{NormalHtLambdaU} for UDW coupling, also hold here. However, investigating the behavior of the QFI versus $ u $, we find that in the low-frequency regime, an increase in $ \omega $ results in the occurrence of the optimal estimation in lower velocities. In particular, as demonstrated in Fig. \ref{NormalHTuOmega}, it does not considerably vary the optimal value of QFI.

\subsection{Practicable measurement for optimal thermometry} \label{practical}
 A major question that may arise is how we can physically implement the optimal thermometry, i.e., a practicable measurement for which the corresponding Fisher information equals the QFI. Noting that the optimal POVM can be made by the eigenvectors of the SLD, we focus on computing them and checking whether these eigenstates overlap with those of some physical observable of the system. Interestingly, following this prescription, we find that when the probe is initially prepared in the ground state ($\theta=\pi$), the optimal POVM can be constructed by the eigenvectors of $ \sigma_{z} $. In other words, the measurement of $ \sigma_{z} $ on the probe leads to the optimal quantum thermometry, saturating the quantum upper bound. This result is absolutely important because not only the optimal POVM but also the maximized QFI is achieved when the atom is initially prepared in the ground state.

	 \subsection{Quantum thermometry in a multiparameter-estimation strategy} \label{multi}
		\begin{figure}[hht]
		\subfigure[]{\includegraphics[width=6cm]{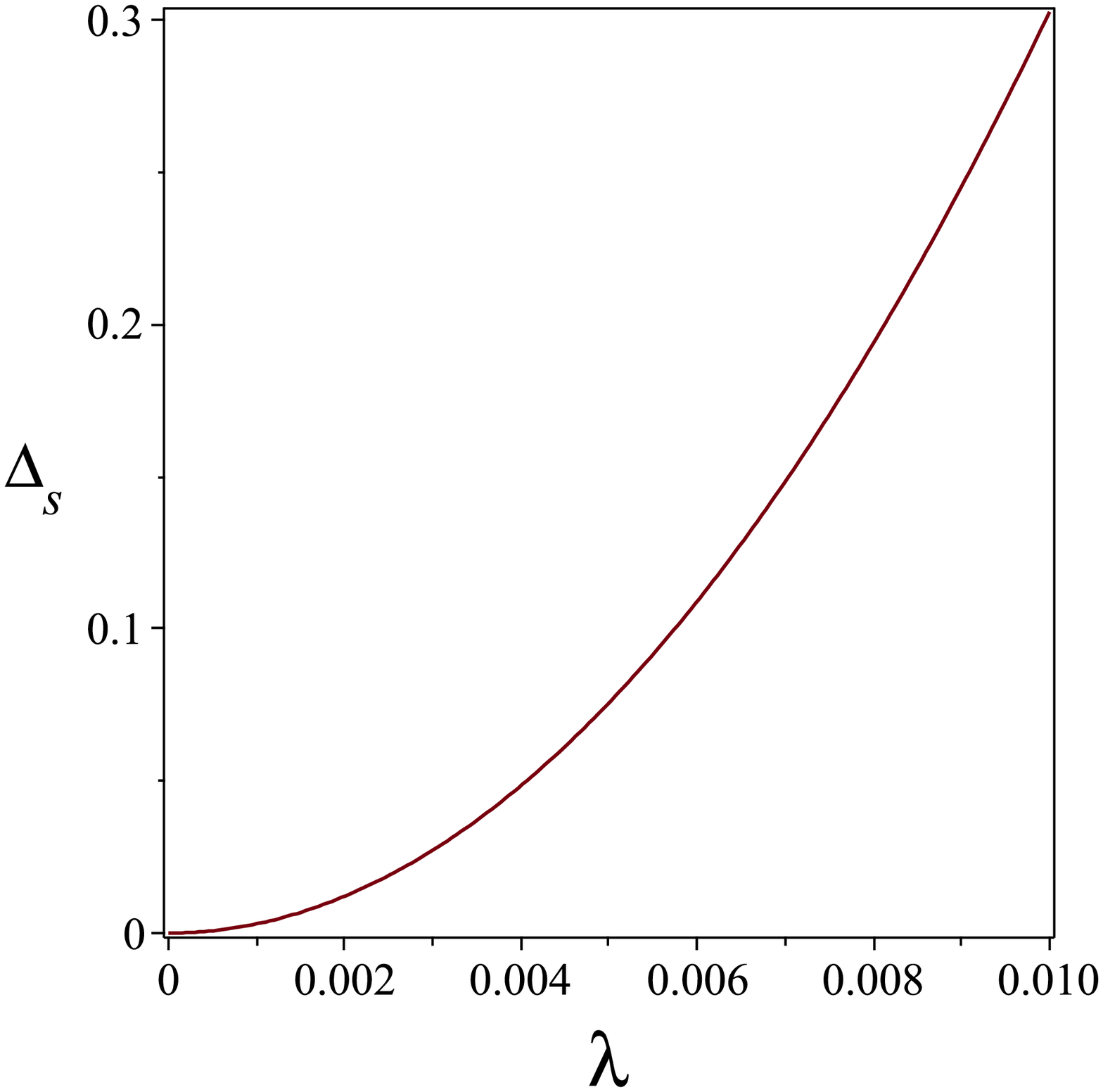}\label{Multi2lambda} }
			\subfigure[]{\includegraphics[width=6cm]{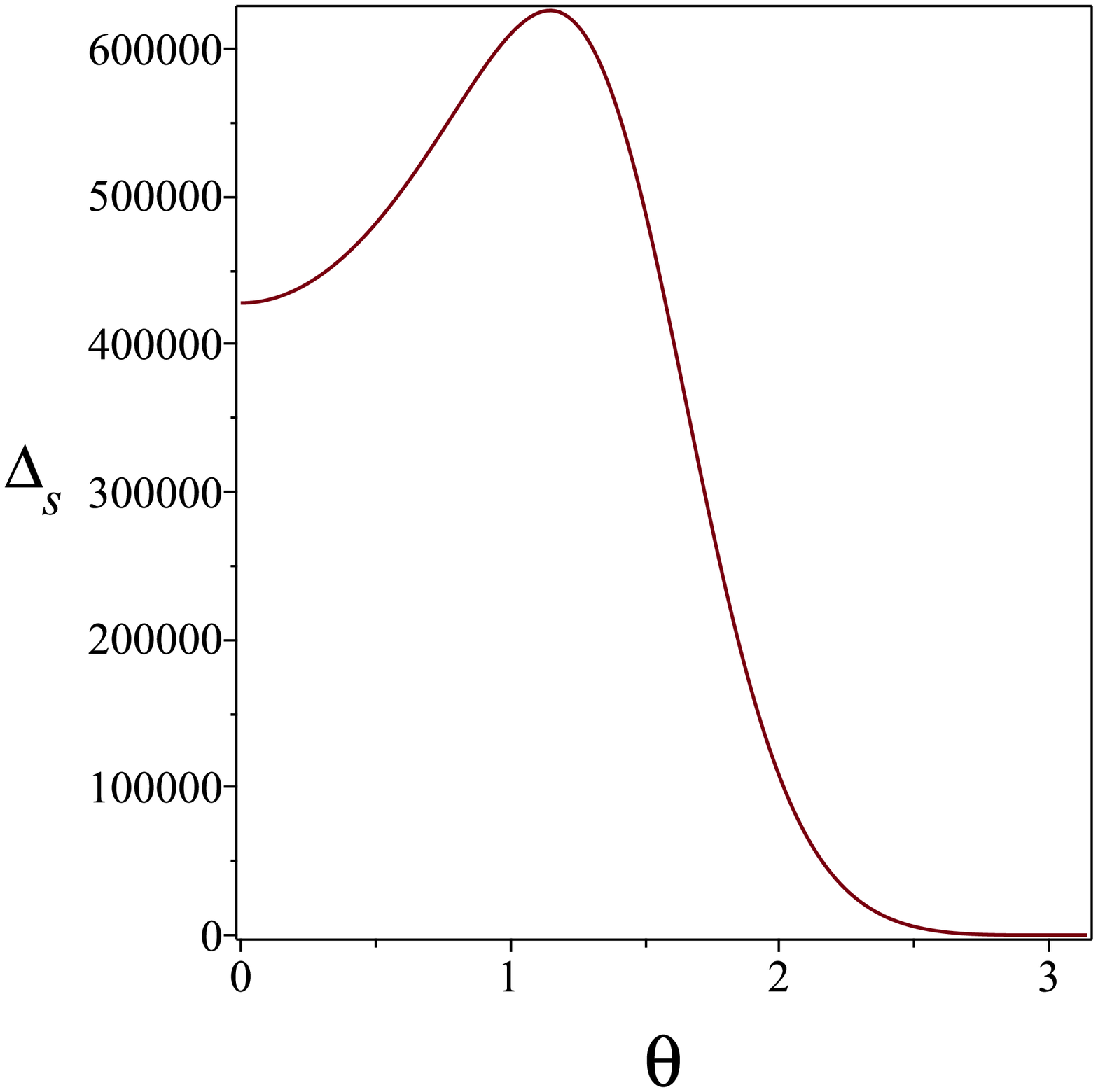}\label{Multi1theta} }
		\caption{UDW-coupling regime: (a) Minimal total variance of simultaneous estimation of $ T $ and $ \theta $ 
			versus   $ \lambda $   for  $ T= 100, \omega=0.1$, $ u=30, \theta=\pi$.
			(b)  The same measure versus   $ \theta $ for   $ T= 0.001, \omega=0.01$, $ \lambda=0.05, u=10$. .}
		\label{Multi1}
	\end{figure}
	
	Because the QFI is usually optimized for $ \theta=\pi $, investigating the simultaneous estimation of $ T $ and $ \theta $ is of great importance for the realization of the optimal thermometry. In particular, we should examine the supremacy of simultaneous estimation in comparison with the individual one. First, some background on the multiparameter estimation theory is presented.

	A quantum system applied in a quantum estimation problem can be characterized by a  quantum state $\rho_\mathbf{\lambda}$ a function of unknown parameters $\mathbf{\lambda}=(\lambda_{1},...,\lambda_{n})$. A multi-parameter quantum estimation strategy is a hunt for the best precision accessible in the simultaneous estimation of $\mathbf{\lambda}$  \cite{szczykulska2016multi}. The quantum Cramer-Rao bound (QCRB), providing a lower bound for the mean square errors of the parameters $\mathbf{\lambda}$, can be formally expressed as \cite{liu2019quantum}
	\begin{equation}\label{Cramerreao}
		\Sigma \geq (M\mathbf{H}_{\lambda})^{-1},
	\end{equation}
	where $ M  $ is the number of experimental runs and $\Sigma=\operatorname{cov}(\hat{\lambda})$ denotes the \textit{covariance matrix }of any locally \textit{unbiased estimators} $\hat{\lambda}$ of the parameters $\lambda$. Moreover,  $\mathbf{H}_{\lambda}$ represents the quantum Fisher information matrix whose components are given by
	\begin{equation}
		(\mathbf{H}_{\lambda})_{i,j}=\frac{1}{2} \operatorname{Tr}\left(\rho_{\lambda}\left\{L_{\lambda_{i}}, L_{\lambda_{j}}\right\}\right)
	\end{equation}
	where $L_{\lambda_{i}}$  denotes the symmetric logarithmic derivatives (SLD),  corresponding to parameter $  \lambda_{i}$, written as
	\begin{equation}
		\frac{L_{\lambda_{i}} \rho_{\lambda}+\rho_{\lambda} L_{\lambda_{i}}}{2}=\partial_{\lambda_{i}} \rho_{\lambda},
	\end{equation}
	in which $\partial_{\lambda_{i}}=\partial / \partial_{\lambda_{i}}$. It should be noted that although the bound in Eq. (\ref{Cramerreao}) is not always tight,  the multiparameter QCRB can be saturated provided that
	the following compatibility condition is satisfied 
	\cite{ragy2016compatibility,napoli2019towards}
	\begin{equation}\label{compatibility}
		\text{Tr}(\rho[L_{\lambda_{i}},L_{\lambda_{k}}])=0.
	\end{equation}
	\par
	Defining ratio \cite{yousefjani2017estimating}:
	\begin{equation}
		R=\frac{\Delta_{\textit{i}}}{\Delta_{\textit{s}}},
	\end{equation}
	where the minimal total variances in the individual and simultaneous estimations are represented, respectively, by $\Delta_{\text{i}}=\sum_{j} \frac{1}{M(\mathbf{H}_{\lambda})_{j,j}}$ and $\Delta_{\textit{s}}=\frac{1}{Mn} \text{Tr}\big(\textbf{H}_{\lambda}^{-1}\big)$, one can collate the performance of the simultaneous estimation in comparison with that of the individual one. Comparing independent and simultaneous schemes, we find that in the simultaneous-estimation strategy fewer resources are required by a factor of the number of parameters to be estimated, and therefore considering $ n $ in the definition of $ \Delta_{\textit{s}} $ is necessary to account for this reduction in resources. The efficiency of simultaneous estimation rather than the independent one can be signified by $R\mathrm{>}$1. Assuming a single run of the experimental measurement, we put $ M=1 $ throughout the paper. 
	
In our model, computing the expectation value of commutator $ [L_{T},L_{\theta}] $ on the probe state, we find that it vanishes, i.e., $ 	\text{Tr}(\rho[L_{T},L_{\theta}])=0 $, indicating that the multiparameter QCRB can be saturated. In other words \cite{vidrighin2014joint}, there is a single measurement that is jointly optimal to extract information on $ T $ and $ \theta $ from the output state, guaranteeing the asymptotic saturability of the QCRB.

In the UDW-coupling regime, more results of interest can be obtained. The most important one is that at a high-velocity regime, $ R $ is roughly maximized, i.e., $ R \approx 2 $,  indicating complete superiority of the simultaneous strategy over the individual one through fast-moving
probes. In general, $ R \leq p $ where $ p $ denotes the number of parameters to be estimated. Moreover, as shown in Fig. \ref{Multi2lambda}, at the high-velocity regime,  $   \Delta_{s} $ grows with an increase in $ \lambda $, and hence better accuracy occurs for weaker couplings. In addition, Fig. \ref{Multi1theta}  demonstrates that total variance $ \Delta_{s}  $ is always minimized for $ \theta=\pi $, indicating the importance of initially preparing the probe in the ground state to achieve the best simultaneous estimation.
 \section{Conclusions}\label{Conclusion}
\par

We investigated relativistic quantum thermometry through a moving probe playing the role of a thermal sensor. It is employed to estimate the temperature of a heat bath modeled by a massless scalar field initially prepared in a thermal state. The effects of the Lamb shift, the initial preparation of the sensor as well as its velocity, and ambient control parameters on the thermometer sensitivity have been analyzed in detail to enhance the quantum estimation. Moreover, quantum thermometry in a multiparameter-estimation scenario has been also addressed.
In addition, the achievement of optimal thermometry and its feasible implementation were precisely discussed. 

An important point which should be addressed is that our results for the thermometry in the low temperature regime may fail for $ T\rightarrow 0 $ (\textit{ultra-low temperatures}, e.g., ion-trap  and cold-atom systems \cite{marzolino2013precision,olf2015thermometry,mehboudi2019using,bouton2020single}) in which thermometry  is timely and challenging.  The reason is that our open quantum system is described by a Markovian master equation, an approximation to
the exact quantum dynamics, in which the nonunitary terms are of second order to the system-environment coupling. The \textit{second order master equation} is derived implementing three approximations \cite{moustos2017non,moustos2018asymptotic}: (i) Born’s approximation applied for weak 
system-environment coupling. 
(ii) The Markov approximation in which the two-time correlation functions of the reservoir are approximated by delta functions.
(iii) The rotating wave approximation
(RWA) ignoring rapidly oscillating terms in the interaction-picture evolution equation \cite{scully1999quantum}. This equation can be utilized in the thermometry when the
sensor-environment interaction is weak and the encoding time
is sufficiently long, where we can regard the sensor as finally evolving to its thermal
equilibrium state independent of the encoding time (i.e., complete thermalization).  
However, sometimes, including  at early times  $ (\tau \thicksim \omega^{-1}) $ \cite{moustos2017non} or  ultra-low temperatures
in which equilibration is slow \cite{moustos2018asymptotic,mitchison2020situ}, and strong system-reservoir couplings when the interaction spectral
density contains zero-value regions \cite{xiong2010exact,cai2014threshold},  the aforementioned  complete-thermalization may be disturbed. 
Particularly, at very low temperatures, the infinitesimal-coupling treatment, relying on local thermalisation of probes, becomes inadequate, because there are quantum correlations between probe and sample, pushing
their marginals far from the Gibbs state \cite{de2016local,correa2017enhancement,miller2018energy,glatthard2022bending}.
In such situations,  the Born-Markov
approximation can provide  analytical results
as well as  intuitionistic pictures, however inevitably might ignore some physical phenomena \cite{wu2021threshold}, and consequently non-Markovian effects are particularly pronounced  \cite{zhang2021non,de2017dynamics}. Although at  later times, the Markov approximation can be usually used safely,  the relaxation
remains non-Markovian for ultra-low temperatures \cite{moustos2017non}.

\par
In addition to the aforementioned points, there is a fundamental limitation for thermometry in \textit{too cold}  samples as $ T/\omega \rightarrow 0 $.  In fact,  the temperature encoded into the probe, at
thermal equilibrium, becomes more difficult to measure the lower it is. In detail, assuming that $\omega $ denotes non-vanishing gap between
 the lowest energy levels of the probe, one can show that for a finite-size quantum probe at equilibrium, the sensing error diverges \textit{ exponentially} as $ T \rightarrow 0 $, known as  Landau bound  \cite{paris2015achieving,correa2015individual,correa2017enhancement,hovhannisyan2018measuring}. It should be noted that
    this ultimate precision   cannot  be purely considered as an intrinsic property of the probe itself. For example, when  the sample is gapless, this bound
     can
    then be surpassed \cite{correa2017enhancement,potts2019fundamental}.
     Indeed,  for  a probe,  strongly coupled
    to a gapless sample,  characterized by a continuous spectrum
    above the ground state, the  thermal sensitivity decays polynomially  (with respect to $ 1/T $),  exhibiting a power-law-like divergence. Similar phenomenon is expected when  the probes are gapless or  are
    not at thermal equilibrium \cite{hovhannisyan2018measuring,potts2019fundamental,planella2022bath,henao2021thermometric}.
      However, it seems that for any total
    system that is not gapless, this exponential divergence cannot be avoided \cite{hovhannisyan2018measuring,potts2019fundamental}. In other words, it has been shown that the key factor when switching between exponential and
    subexponentially inefficient quantum thermometry is whether
    the energy spectrum of the global many-body system
    exhibits a finite gap   or not \cite{hovhannisyan2018measuring}.

    \par
    
    In many potential applications of quantum estimation theory, the region in which we have to probe is out of our reach, or the sensor should monitor the entire area to gain complete information. Moreover, the metrological devices may be located or accessible at another place. In these cases, employing moving sensors is of key importance. Therefore, a more rigorous investigation of probing environments using moving sensors is required. In particular, the idea can be generalized to situations in which two or more entangled sensors are applied to enhance quantum estimation.

\section*{Declaration of competing interest}
The authors declare that they have no competing interests.

\section*{Acknowledgements}
H.R.J. wishes to acknowledge the financial support of the MSRT of Iran and Jahrom University. 
R.L.F. acknowledges support from Unione Europea -- NextGenerationEU -- fondi MUR D.M. 737/2021 -- progetto di ricerca ``IRISQ''.

\bibliography{Ref}

\end{document}